\documentclass[footinbib,twocolumn,showpacs,amsmath,amstex,amssymb,mathfonts,superscriptaddress,prx]{revtex4}
\usepackage{graphicx}
\usepackage{color}
\usepackage{bm}

\usepackage{graphicx}
\usepackage{color}
\usepackage{bm}
\usepackage{amsmath}
\usepackage{amssymb}
\usepackage{amsthm}
\usepackage{amsfonts}
\usepackage{multirow}
\usepackage{makecell}
\usepackage{float}
\usepackage{comment}
\usepackage{enumitem}
\usepackage{verbatim}
\usepackage{mathtools}
\usepackage{esint}
\usepackage{tipa}

\usepackage{bbm}

\begin{document}

\title{Microscopic theory for nematic fractional quantum Hall effect}
\author{Bo Yang} 
\affiliation{Division of Physics and Applied Physics, Nanyang Technological University, Singapore 637371.}
\affiliation{Institute of High Performance Computing, A*STAR, Singapore, 138632.}
\pacs{73.43.Lp, 71.10.Pm}

\date{\today}
\begin{abstract}
We analyse various microscopic properties of the nematic fractional quantum Hall effect (FQHN) in the thermodynamic limit, and present necessary conditions required of the microscopic Hamiltonians for the nematic FQHE to be robust. Analytical expressions for the degenerate ground state manifold, ground state energies, and gapless nematic modes are given in compact forms with the input interaction and the corresponding ground state structure factors. We relate the long wavelength limit of the neutral excitations to the guiding center metric deformation, and show explicitly the family of trial wavefunctions for the nematic modes with spatially varying nematic order near the quantum critical point. For short range interactions, the dynamics of the FQHN is completely determined by the long wavelength part of the ground state structure factor. The special case of the FQHN at $\nu=1/3$ is discussed with new theoretical insights from the Haffnian parent Hamiltonian, leading to a number of rigorous statements and experimental implications.
\end{abstract}

\maketitle 

\section{Introduction}

For condensed matter systems with non-trivial topological orders, the robustness of the topological properties at low temperature usually requires the ground state to have a finite energy gap to all excitations in the thermodynamic limit\cite{prange}. In general for such systems, the universal topological features dominate the ground state response, and the geometric properties of the system are less important. The incompressible FQH states are such examples where topological orders arise from strong interactions between electrons, without needing protection of any symmetry. There are also examples of compressible FQH states with no plateau formation for the Hall conductivity, with anisotropic stripe or bubble phases that are gapless and spontaneously break the rotational/translational symmetry\cite{west1,west2,west3,west4,klitzing,lin,csathy,smet,manfra}. An interesting exception is the nematic fractional quantum Hall effect (FQHN), which was recently discovered in experiments\cite{xia,west5}. Here we have examples where topological orders and non-trivial geometric effects coexist: there is an anisotropic longitudinal resistivity enhanced by low temperature, and at the same time with a robust plateau for Hall conductivity.

It is generally believed that the robustness of the Hall conductivity plateau in FQHN is due to the finite charge gap, while the anisotropic longitudinal resistivity is a result of the neutral excitations in the long wavelength limit becoming gapless\cite{joseph}. Such neutral excitations form a degenerate ground state manifold. They are thus prone to spontaneous symmetry breaking. It is well known that the neutral excitations in the long wavelength limit is a quadrupole excitation that breaks rotational symmetry, potentially leading to anisotropic transport\cite{yangbo1,regnault}. Non-trivial geometric effects also arise in experiments where rotational symmetry is explicitly broken\cite{mansour1,mansour2}. Several field theoretical studies of the FQHN have been carried out, either by assuming that the neutral excitations go soft in the long wavelength limit\cite{joseph}, or by adding an attractive quadrupolar interaction\cite{yizhi,rezayi}. These theories capture the topological order, the nematic order from spontaneous symmetry breaking, as well as the neutral and charge gaps\cite{gromov1,gromov2,gromov3} in a phenomenological manner.

Microscopic theories are needed to better understand the assumptions used in the field theoretical approaches. For the FQHN most studies so far are based on numerical computations. Finite system analysis has established that the single mode approximation (SMA) is exact for the neutral excitations in the long wavelength limit\cite{yangbo1}. This is true from numerical calculations for all accessible system sizes, and is expected to be true in the thermodynamic limit. The Jack polynomial formalism, the composite fermion picture and the first quantised form of the neutral excitations are also constructed to shed more insights on the nature of such many-body states\cite{yangbo1,sreejith,ajit1,ajit2,yangbo2,gromov}. Numerical studies have tentatively shown that short range interactions can lead to instability of the intrinsic guiding center metric, and such ``squeezed" Laughlin states can harbour uniform nematic order\cite{regnault}. It is, however, difficult to show microscopically how assumptions in the FQHN field theory can arise from bare interactions between electrons with numerical studies. In particular, important physics happening at long wavelength limit is inaccessible given the relatively small system sizes that can be computed numerically.

In this paper, we compute analytically the conditions for the long wavelength limit (small $q$) of the neutral excitation to go soft in the thermodynamic limit. Using the Laguerre polynomials as the basis, variational energies of the neutral excitations at small $q$ is controlled by two universal, tridiagonal characteristic matrix $\Gamma_{\left(1\right)},\Gamma_{\left(2\right)}$ that can be computed exactly and are independent of microscopic details. The SMA at small $q$ becomes exact eigenstates when it is degenerate with the ground state, and we can identify it with the guiding center metric deformation of the ground state. Thus the onset of the FQHN can be understood as the case when the shear modulus of the gapped ground state of the quantum fluid vanishes\cite{haldane3}. 

We also identify trial wavefunctions for the gapless nematic mode from spontaneous symmetry breaking, where the spatially varying ``nematic wave" can be shown explicitly. While $\Gamma_{\left(1\right)}$ controls the neutral excitation gap, the dispersion of the nematic mode is controlled by $\Gamma_{\left(2\right)}$. The tridiagonal nature of $\Gamma_{\left(1\right)}$ and $\Gamma_{\left(2\right)}$ implies the dynamics of the FQHN only depends on the long wavelength part of the ground state structure factor, if the interaction is short-ranged. The analysis here can much simplify the numerical computation of the nematic phase and its finite size scaling. The derived results are applicable to FQH phases \emph{at any filling factor}. The necessary analytic conditions for the robustness of the FQHN phases are also illustrated with numerical calculations using the Laughlin state at filling factor $\nu=1/3$ as an example. 

The FQHN phase at $\nu=n+1/3$ is also special, because we can show that the quadrupole excitations are exact zero modes of the Haffnian model Hamiltonian. We will thus use it as an example to illustrate the validity of the general methodologies (both analytic and numerical) proposed in this paper. The connection to the Haffnian state also allows us to derive a family of two-body interactions supporting robust FQHN in the presence of strong magnetic field, which can be realised experimentally with suitable tuning of the sample thickness and interaction screening. We also show the presence of Landau level (LL) mixing can potentially help stabilise FQHN in higher LLs, pointing to diverse conditions for the experimental observations of the FQHN in the neighbourhood of the fully gapped Laughlin phase.

This paper will be organised as follows: In Sec.~\ref{sma} we compute the long wavelength energy gap of the neutral excitations from the SMA in the thermodynamic limit, and show that it is determined by the universal characteristic matrix $\Gamma_{\left(1\right)}$. We term such neutral excitations in the long wavelength limit as the quadrupole excitations. In Sec.~\ref{shear} we show the quadrupole excitations can be identified as a uniform area-preserving deformation of the ground state, both from the wavefunction and the energetics perspectives. Thus the quadrupole excitations harbour uniform nematic order\cite{regnault}. In Sec.~\ref{goldstone} we derive the expression of the spatially varying nematic order from the trial wavefunctions of the gapless nematic mode in the FQHN phase. We also show the quadratic dispersion of the nematic mode is controlled by another universal characteristic matrix $\Gamma_{\left(2\right)}$. In Sec.~\ref{mm} we analytically investigate several families of short range microscopic models, and derive conditions for the FQHN to be viable. In Sec.~\ref{ms} we carry out preliminary numerical analysis focusing on the Laughlin phase at filling factor $\nu=1/3$, corroborating with the analytical results to show tentative evidence of FQHN when the two-body interaction is a family of linear combinations of the $\hat V_1^{\text{2bdy}}, \hat V_3^{\text{2bdy}}$ and $\hat V_5^{\text{2bdy}}$ pseudopotentials derived from eigenstates of $\Gamma_{\left(1\right)}$. In Sec.~\ref{qd} we discuss about the contrasting natures of the quadrupole and dipole neutral excitations at $\nu=n+1/3$, showing the interesting connection of the quadrupole excitations to the Haffnian model Hamiltonian, with various experimental implications. In Sec.~\ref{con} we summarize the results of this paper and discuss about the future works.

\section{SMA in the long wavelength limit}\label{sma}
Let us start with a two-body Hamiltonian in a single Landau level (LL) as follows:
\begin{eqnarray}\label{h}
\hat{\mathcal H}=\int \frac{d^2q}{4\pi}V_{\bm q}\hat\rho_{\bm q}\hat\rho_{-\bm q}
\end{eqnarray}
where $\hat\rho_{\bm q}=\sum_ie^{iq_a\hat R_i^a}$ is the guiding center density operator, and $\hat R^a_i$ are the guiding center coordinates with only matrix elements between states in the same Landau level. It also satisfies the commutation relation $[\hat R_i^a, \hat R_i^b]=-i\epsilon^{ab}l_B^2$, where $l_B$ is the magnetic length. The number of electrons is given by $N_e$ and we set $l_B=1$. Assuming at a fixed filling factor, Eq.(\ref{h}) is incompressible with both neutral and charged quasielectron gaps, with ground state $|\psi_0\rangle$ and energy $E_0$. Defining the regularised guiding center density as $\delta\hat\rho_{\bm q}=\hat\rho_{\bm q}-\langle\psi_0|\hat\rho_{\bm q}|\psi_0\rangle=\hat\rho_{\bm q}-\frac{N_e\delta\left(q\right)}{2\pi q}$ with $q=|\bm q|$, the GMP algebra\cite{gmp} is given by:
\begin{eqnarray}
[\delta\hat\rho_{\bm q_1},\delta\hat\rho_{\bm q_2}]=2i\sin\frac{\bm q_1\times\bm q_2}{2}\delta\hat\rho_{\bm q_1+\bm q_2}
\end{eqnarray}
The regularised ground state structure factor is defined as $S_{\bm q}=\langle\psi_0|\delta\hat\rho_{\bm q}\delta\hat\rho_{-\bm q}|\psi_0\rangle$ and we also have the following relationship for fermions\cite{haldane1,haldane2}:
\begin{eqnarray}\label{fourier}
s_{\bm q}=S_{\bm q}-S_\infty=-\int\frac{d^2q'}{2\pi}e^{i\bm q\times \bm q'}s_{\bm q'}
\end{eqnarray}
We now start with the family of SMA trial wavefunctions $|\psi_{\bm q}\rangle=\delta\hat\rho_{\bm q}|\psi_0\rangle$, which are orthogonal to the ground state with variational energies $E_{\bm q}$. The variational energy gap is thus given by\cite{sup}:
\begin{eqnarray}
&&\delta E_{\bm q}=\frac{\langle\psi_{\bm q}|\hat{\mathcal H}|\psi_{\bm q}\rangle}{\langle\psi_{\bm q}|\psi_{\bm q}\rangle}-E_0=\frac{\langle\psi_0|[\delta\hat\rho_{-\bm q},[\hat{\mathcal H},\delta\hat\rho_{\bm q}]]|\psi_0\rangle}{2S_{\bm q}}\qquad\\
&&=\frac{1}{2S_{\bm q}}\int \frac{d^2q'}{4\pi}V_{\bm q'}\left(2\sin\left(\frac{1}{2}\bm q'\times\bm q\right)\right)^2\nonumber\\
&&\qquad\qquad\times\left(s_{\bm q'+\bm q}+s_{\bm q'-\bm q}-2s_{\bm q'}\right)\label{sss}
\end{eqnarray}
Here we assume rotational invariance. In the long wavelength limit, the ground state structure factor is given by $\lim_{|q|\rightarrow 0}S_{\bm q}=\eta q^4$, where $\eta=N_e\kappa/2$ and $\kappa$ is bounded below by the Hall viscosity of the ground state\cite{haldane3}. By expanding Eq.(\ref{sss}) to the leading order in $q$, and using Eq.(\ref{fourier}), we have the following expression:
\begin{eqnarray}\label{eq0}
&&\delta E_{\bm q\rightarrow 0}=\frac{1}{2\eta}\int \frac{d^2q'd^2q''}{8\pi^2}V_{\bm q'}\left(q'_xq''_x\right)^2e^{i\bm q'\times\bm q''}s_{\bm q''}+O\left(q^2\right)\nonumber\\
&&=\frac{1}{512\eta}\iint_0^\infty dq_1dq_2V_{q_1}s_{q_2}q_1q_2\nonumber\\
&&\times\left(\prescript{}{0}{\mathbf{F}}_1\left(1,-\frac{q_1q_2}{4}\right)+\prescript{}{0}{\mathbf{F}}_1\left(2,-\frac{q_1q_2}{4}\right)\right)+O\left(q^2\right)\qquad
\end{eqnarray}
Here $q_1=|q'|^2,q_2=|q''|^2$, and $\prescript{}{0}{\mathbf{F}}_1\left(a,x\right)$ is the regularised hypergeometric function\cite{sup}. For very short range interactions (e.g. with $
\hat V_1^{\text{2bdy}}$ pseudopotential\cite{prange}), $\delta E_{\bm q\rightarrow 0}>0$ and is buried in the continuum of multi-roton excitations. If $V_{\bm q}$ in Eq.(\ref{h}) can be tuned such that $\delta E_{\bm q\rightarrow 0}\rightarrow 0$, then $|\psi_{\bm q}\rangle$ becomes an exact eigenstate, degenerate with $|\psi_0\rangle$, given that there is no level crossing from $\hat V_1^{\text{2bdy}}\rightarrow V_{\bm q}$.

To evaluate the numerator in Eq.(\ref{eq0}), we first note that due to the property of the structure factor in Eq.(\ref{fourier}), $s_{\bm q}$ is a linear combination of Laguerre polynomials $L_{m}\left(q^2\right)$ with odd $m$. Expanding $V_{\bm q}$ in the same basis of Laguerre polynomials, we have the following:
\begin{eqnarray}\label{eq0a}
&&\delta E_{\bm q\rightarrow 0}=\frac{1}{256\eta}\Gamma^{mn}_{\left(1\right)}c_md_n+O\left(q^2\right)\\
&&V_{\bm q}=\sum_{m}c_me^{-\frac{q^2}{2}}L_m\left(q^2\right),s_{\bm q}=\sum_{n}d_ne^{-\frac{q^2}{2}}L_n\left(q^2\right)\qquad\\
&&\Gamma^{mn}_{\left(1\right)}=\frac{1}{2}\iint_0^\infty dq_1dq_2e^{-\frac{q_1+q_2}{2}}q_1q_2L_m\left(q_1\right)L_n\left(q_2\right)\nonumber\\
&&\qquad\quad\times\left(\prescript{}{0}{\mathbf{F}}_1\left(1,-\frac{q_1q_2}{4}\right)+\prescript{}{0}{\mathbf{F}}_1\left(2,-\frac{q_1q_2}{4}\right)\right)\label{gammamn}
\end{eqnarray}
Note the two-body Haldane pseudopotential interaction Hamiltonians are also given by the Laguerre polynomials:
\begin{eqnarray}
\hat V_n^{\text{2bdy}}=\int \frac{d^2q}{4\pi}e^{-\frac{q^2}{2}}L_n\left(q^2\right)\hat\rho_{\bm q}\hat\rho_{-\bm q}
\end{eqnarray}
Using the Hardy–Hille formula, Eq.(\ref{gammamn}) can be further simplified to give:
\begin{eqnarray}\label{gamma1}
\Gamma^{mn}_{\left(1\right)}&=&[(1-m)m\delta_{m,2+n}-(1+m) (2+m)\delta_{m,n-2}\nonumber\\
&&+2(1+m+m^2)\delta_{m,n}]\left(-1\right)^m
\end{eqnarray}
where both $m,n$ are odd integers, and $\Gamma^{mn}_{\left(1\right)}$ is a tridiagonal matrix. It is then useful to treat $V_{\bm q}, s_{\bm q}$ as vectors $\bm c, \bm d$ respectively, in the basis of Laguerre polynomials, where $\bm d$ is completely from the ground state. The \emph{dot product} $\bm c\cdot\bm d$ gives the ground state energy $E_0$, and the variational energy gap is given by the inner product: 
\begin{eqnarray}\label{compact1}
\delta E_{\bm q\rightarrow 0}=\frac{1}{256\eta}\langle\bm c,\bm d\rangle_{\Gamma_{\left(1\right)}}+O\left(q^2\right).
\end{eqnarray}

Note that $\Gamma_{\left(1\right)}$ is a well-defined mathematical function given by Eq.(\ref{gamma1}), while the only physical input to the Hamiltonian is given by $\bm c$. There is a one-to-one mapping of $\bm d$ from $\bm c$, with the ground state of Eq.(\ref{h}). For short range interactions with $c_m=0$ for $m>m_0$, we only need to consider $d_n$ with $n\le m_0+2$. A more detailed analysis will be presented in Sec.~\ref{mm}.

\section{Nematic order for the neutral excitations}\label{shear}
We now explore the nematic order of the neutral excitations in the long wavelength limit by connecting them to the anisotropic ground state from deforming the guiding center metric of $|\psi_0\rangle$. The area-preserving deformation generators can be defined as $\hat\Lambda^{ab}=\frac{1}{4}\sum_i\{\hat R_i^a,\hat R_i^b\}$ with the following closed algebra\cite{haldane3}:
\begin{eqnarray}
[\hat\Lambda^{ab},\hat\Lambda^{cd}]=\frac{i}{2}\left(\epsilon^{ac}\hat\Lambda^{bd}+\epsilon^{ad}\hat\Lambda^{bc}+\epsilon^{bc}\hat\Lambda^{ad}+\epsilon^{bd}\hat\Lambda^{ac}\right)\quad\quad
\end{eqnarray}
The family of anisotropic ground states can thus be defined as $|\xi_{\theta,\phi}\rangle=e^{i\alpha_{ab}\hat\Lambda^{ab}}|\psi_0\rangle$, with $\alpha_{ab}$ as a symmetric matrix. The Bogoliubov transformation of the guiding center coordinates is given by $\hat R'^a=\lambda_b^a\hat R^b=e^{-i\alpha_{cd}\hat\Lambda^{cd}}\hat R^ae^{i\alpha_{cd}\hat\Lambda^{cd}}$, thus $|\xi_{\theta,\phi}\rangle$ is the ground state of Eq.(\ref{h}) with the transformation in $V_{\bm q}$: $q_a\rightarrow \left(\lambda^{-1}\right)_a^bq_b$, or $q^2\rightarrow g^{ab}q_aq_b$, where $g^{ab}$ is a unimodular metric parametrised as follows:
\begin{eqnarray}\label{matrix}
g=\left(\begin{array}{cc}
\cosh\theta+\sinh\theta\cos\phi & \sinh\theta\sin\phi\\
\sinh\theta\sin\phi & \cosh\theta-\sinh\theta\cos\phi\end{array}\right)
\end{eqnarray}
For the rotationally invariant $|\psi_0\rangle$, the variational energy of $|\xi_{\theta,\phi}\rangle$ only depends on $\theta$, which parameterises the squeezing of the metric, as follows\cite{sup}:
\begin{eqnarray}\label{compact2}
&&\lim_{\theta\rightarrow 0}\delta E_\alpha=\langle\xi_{\theta,\phi}|\hat{\mathcal H}|\xi_{\theta,\phi}\rangle-E_0=\frac{1}{64}\langle\bm c,\bm d\rangle_{\Gamma_{\left(1\right)}}\theta^2\qquad
\end{eqnarray}
Comparing Eq.(\ref{compact1}) and Eq.(\ref{compact2}), we can see the variational energy of the neutral excitations in the long wavelength limit is related to the shear modulous $\langle\bm c,\bm d\rangle_{\Gamma_{\left(1\right)}}$ of the ground state. Thus for small $|\bm q|$ and $\theta$, $|\psi_{\bm q}\rangle$ and $|\xi_{\theta,\phi}\rangle$ approximately have the same energy with $\theta=1/\left(2\sqrt \eta\right)$.

In general $|\psi_{\bm q\rightarrow 0}\rangle$ and $|\xi_{\theta\rightarrow 0,\phi}\rangle$ do not have to be related to each other even when they have the same variational energy. However when $\langle\bm c,\bm d\rangle_{\Gamma_{\left(1\right)}}\rightarrow 0$, they will belong to the same manifold of degenerate ground states. Denoting $\phi_{\bm q}$ as the angle of the momentum, we can identify the following at small $q$ based on inversion symmetry, as long as $|\psi_{\bm q}\rangle$ are the only states degenerate with the ground state:
\begin{eqnarray}\label{nematicg}
|\xi_{\theta,\phi_{\bm q}}\rangle\sim|\psi^{\pm}_{\bm q}\rangle=\frac{1}{\sqrt{2S_{\bm q}}}\left(|\psi_{\bm q}\rangle\pm|\psi_{-\bm q}\rangle\right)
\end{eqnarray}
This is the ground state of the FQHN after spontaneous symmetry breaking, and finite size numerical analysis indicates that $|\xi_{\theta,\phi}\rangle$ could have uniform nematic order\cite{regnault}. Thus in the long wavelength limit, $|\psi_{\bm q}^{\pm}\rangle$ is equivalent to the guiding center metric deformation of the ground state at $\bm q=0$, if the shear modulus $\langle\bm c,\bm d\rangle_{\Gamma_{\left(1\right)}}$ vanishes. This leads to the development of the nematic order for the neutral excitations in this limit.

\section{Gapless Nematic Wave at Critical Point}\label{goldstone}
The long wavelength spatial modulation of the nematic order should gives rise to the gapless excitations that are related to the Goldstone mode. To identify these states let us first define the operator of the local nematic order, which is a slightly modified version from\cite{regnault}:
\begin{eqnarray}\label{nematic}
\hat{\mathcal N}=\int_0^{2\pi}\frac{d\theta_{\bm l}}{2\pi}e^{2i\theta_{\bm l}}\lim_{|\bm l|\rightarrow 0}\frac{1}{|\bm l|^2}\delta\hat\rho\left(\bm r+\bm l/2\right)\delta\hat\rho\left(\bm r-\bm l/2\right)\quad
\end{eqnarray}
where $\bm l$ is an arbitrary point-splitting vector, $\theta_{\bm l}$ is the angle of $\bm l$, and $\delta\hat\rho\left(\bm r\right)$ is the Fourier component of $\delta\hat\rho_{\bm q}$. For a translationally invariant state $|\psi_0\rangle$, the nematic order is independent of $\bm r$, and we have:
\begin{eqnarray}\label{nematic1}
\langle\psi_0|\hat{\mathcal N}|\psi_0\rangle=-\lim_{|\bm l|\rightarrow 0}\frac{1}{|\bm l|^2}\int_0^{2\pi}\frac{d\theta_{\bm l}}{2\pi}e^{2i\theta}s_{\tilde{\bm l}}
\end{eqnarray}
where $s_{\tilde{\bm l}}$ is defined by Eq.(\ref{fourier}) with $\tilde{l}_a=l_B^{-2}\epsilon_{ab}l^b$, $l_B$ being the magnetic length. Eq.(\ref{nematic1}) is clearly zero if $|\psi_0\rangle$ is rotationally invariant, and non-zero if the structure factor has a quadrupole symmetry. For the nematic ground state established in Eq.(\ref{nematicg}), simple algebra gives us:
\begin{eqnarray}\label{nematic2}
\langle\psi^{\pm}_{\bm q}|\hat{\mathcal N}|\psi^{\pm}_{\bm q}\rangle=\mathcal N_{q}^{\left(1\right)}\pm\cos 2qr\mathcal N_{q}^{\left(2\right)}
\end{eqnarray}
where $\mathcal N_{q}^{\left(1\right)},\mathcal N_{q}^{\left(2\right)}$ are two non-universal functions of $q$ that can be computed analytically\cite{sup}. Thus at least when $q$ is small enough, $|\psi^{\pm}_{\bm q}\rangle$ is the nematic mode with spatially varying nematic order given by the second term in Eq.(\ref{nematic2}).

To look at the dispersion of this nematic wave, we can expand Eq.(\ref{sss}) to the next order. When Eq.(\ref{compact1}) vanishes at the FQHN phase, we have\cite{sup}:
\begin{eqnarray}
\delta E_{\bm q\rightarrow 0}&=&\frac{1}{768\eta}\langle\bm c,\bm d\rangle_{\Gamma_{\left(2\right)}}q^2+O(q^4)\label{gamma2}\\
\Gamma_{\left(2\right)}^{mn}&=&\left(-1\right)^m[(2m-1)(m-1)m\delta_{m,2+n}\nonumber\\
&&+(1+m) (2+m)(2m+3)\delta_{m,n-2}\nonumber\\
&&-2(1+2m)(1+m+m^2)\delta_{m,n}]
\end{eqnarray}
It is important to note that the dispersion of the gapless nematic mode is quadratic. The necessary condition for the FQHN phase is thus $\langle\bm c,\bm d\rangle_{\Gamma_{\left(1\right)}}=0,\langle\bm c,\bm d\rangle_{\Gamma_{\left(2\right)}}>0$. Both $\Gamma_{\left(1\right)}$ and $\Gamma_{\left(2\right)}$ are universal tridiagonal matrices independent of the microscopic details of the Hamiltonians.

Effective field theories generally predicts a linear gapless Goldstone mode in the FQHN phase, from the long wavelength fluctuation of the nematic director\cite{joseph,yizhi,gromov1}. The velocity of this Goldstone mode vanishes at the quantum critical point (QCP) when the neutral mode becomes degenerate with the ground state, leading to a quadratic dispersion from the nematic amplitude fluctuation. The microscopic theory agrees with this effective description in the neighbourhood of the QCP with $\langle\bm c,\bm d\rangle_{\Gamma_{\left(1\right)}}\sim 0$, where the analytical derivation of the variational energies and the quantum states are exact. Deep in the isotropic phase where $\langle\bm c,\bm d\rangle_{\Gamma_{\left(1\right)}}>0$, our calculations will only be accurate if the SMA still gives good variational wavefunctions of these Hamiltonians.

It is important to highlight that deep in the nematic phase when $\langle\bm c,\bm d\rangle_{\Gamma_{\left(1\right)}}<0$ and the global ground state is no longer isotropic, the long wavelength SMA  may not be good trial wavefunctions. This could lead to the microscopic theory here not able to capture the linear Goldstone mode, and it does not preclude the existence of such linear dispersions away from the critical point. On the other hand, moving deep into the nematic phase generally implies moving further away from the model Hamiltonians of the fully gapped FQH states, which will likely close the charge gap and destroy the FQHN phase. Gapless smectic or stripe phases are expected especially in higher LLs. It is also worth noting that unlike quantum Hall ferromagnets, the nematic director in FQHN are not really physically measurable quantities. Both the fluctuations of the nematic amplitude and the nematic director lead to the fluctuation of the nematic order defined in Eq.(\ref{nematic}), associated with the quadratic gapless dispersion. Further investigations are warranted both for the effective and the microscopic theories in the nematic phase far away from the QCP.  

\section{Minimal models for nematic fractional quantum Hall effect}\label{mm}

To understand the dynamics of the FQHN phase from microscopic Hamiltonians analytically as much as possible, we start with spectrum of $\Gamma_{\left(1\right)}$, which is real given that the matrix is symmetric. The eigenvalues $\lambda_1$ and corresponding eigenvectors $\vec c^{\lambda_1}$ satisfy the following relationship:
\begin{eqnarray}\label{eigenrelation}
c^{\lambda_1}_{k+2}=\frac{\left(\lambda-\Gamma^{k,k}_{\left(1\right)}\right)c^{\lambda_1}_{k}+\Gamma^{k,k-2}_{\left(1\right)}c^{\lambda_1}_{k-2}}{\Gamma^{k,k+2}_{\left(1\right)}}\quad
\end{eqnarray}
where $k$ is a non-negative odd integer and $c^{\lambda_1}_{-1}=0$. In particular, if the microscopic two-body interaction is $\vec c=\vec c^{\lambda_1}$, we then have $\lim_{\bm q\rightarrow 0}\delta E_{\bm q}=\left(\lambda_1 E_0\right)/\left(256\eta\right)$, where $E_0=\vec c\cdot\vec d$ is the ground state energy in the $\bm q=0$ sector. It is easy to check that $\lambda_1=0$ gives $c^0_{k}=\textit{const}$, which is not relevant for realistic interactions. 

We now focus on a special family of interactions with $\vec c$ such that $c_k=c^{\lambda_1}_k$ for $k\le k_0$, and $c_k=0$ for $k>k_0$. These are interactions from the eigenvectors of $\Gamma_{\left(1\right)}$ but with a cut-off. For the more realistic case where $c_{k}$ decreases with $k$, we need to have $\lambda_1<0$. Simple algebra leads to:
\footnotesize
\begin{eqnarray}\label{master1}
\delta E_{\bm q}=\frac{\left(\lambda_1 E_0+\Gamma^{k_0,k_0+2}_{\left(1\right)}\left(c^{\lambda_1}_{k_0}d_{k_0+2}-c^{\lambda_1}_{k_0+2}d_{k_0}\right)\right)}{256\eta}+O(q^2)\quad
\end{eqnarray}
\normalsize
Thus the variational energy gap requires three inputs from numerical computations: $d_{k_0},d_{k_0+2}$ and the ground state energy $E_0$. This relationship is valid at any filling factor in the thermodynamic limit. 

\subsection{$k_0=1$}
Without loss of generality, we always set $c_1=c_1^{\lambda_1}=1$. The simplest case is for $k_0=1$. For the Laughlin state at filling factor $\nu=1/3$, it is the model Hamiltonian leading to $E_0=d_1=0$. This gives us:
\begin{eqnarray}
\delta E_{\bm q}=\frac{3}{128\eta}d_3-\frac{15}{128\eta}d_3q^2+O(q^4)
\end{eqnarray}
from Eq.(\ref{master1}) and Eq.(\ref{gamma2}). The neutral mode is gapped with a negative dispersion at $\bm q\rightarrow 0$, as it should be. This is also true for filling factor $\nu\le1/3$. More precisely, let $N_e, N_o$ be the number of electrons and number of orbitals respectively on the sphere or disk geometry, we then have $d_1=0$ for $N_o>3N_e-2$. 

For $N_o<3N_e-2$, $d_1$ does not vanish, and the variational energy gap is given by:
\begin{eqnarray}\label{v1}
\delta\tilde E_{\bm q}=\frac{3}{128\eta}\left(d_3-d_1\right)+\frac{3}{384\eta}\left(3d_1-5d_3\right)q^2+O(q^4)\quad
\end{eqnarray}
Thus for very short-range interactions (in the neighbourhood of pure $\hat V_1^{\text{2bdy}}$ pseudopotential), the necessary condition for a gapped translationally invariant ground state is for $d_3>d_1$. The global ground state will no longer be translationally invariant with $d_1\ge d_3$, and spontaneous symmetry breaking will generally occur. For $d_1>5d_3/3$, the dispersion of the neutral excitation is positive, indicating a possibility of the charged gap and quantised Hall conductivity. We will explore these possibilities in Sec.~\ref{ms}. 

\subsection{$k_0=3$}
For the case of $k_0=3$, the model Hamiltonian is a linear combination of the $\hat V_1^{\text{2bdy}},\hat V_3^{\text{2bdy}}$ pseudopotentials (with coefficients $c_1,c_3$). This can be fully tuned by $\lambda_1$, with $c_3=1+\lambda_1/6$. The physically relevant regime is thus for $c_3>0$ and $\lambda_1>-6$. Let the eigenvectors of $\Gamma_{\left(2\right)}$ be $\vec c^{\lambda_2}$ with eigenvalue $\lambda_2$, the following expression can be obtained with some algebraic manipulation:
\begin{eqnarray}\label{k03}
&&\delta E_{\bm q}\propto \lambda_1 E_0+20\left(c_3d_5-c^{\lambda_1}_5d_3\right)\nonumber\\
&&+\left(\lambda_2 E_0-180\left(c_3d_5-c_5^{\lambda_2}d_3\right)\right)q^2+O(q^4)
\end{eqnarray}
Since we are only interested in the signs of each term, we ignore the denominator in Eq.(\ref{k03}). We also have $\lambda_1=6c_3-6, \lambda_2=18-30c_3,c^{\lambda_1}_5=1+4\lambda_1/15+\lambda_1^2/120, c_5^{\lambda_2}=11/25-\lambda_2/27+\lambda_2^2/5400$. Given that $E_0=\vec c\cdot \vec d>c_3d_3$, the condition for the first line of Eq.(\ref{k03}) to be zero, and the coefficient of the second line to be positive, leads to a narrow range in the parameter space of $c_3$ and $d_5/d_3$ as shown in Fig.(\ref{fig2}). Note that $d_5/d_3$ is not an independent parameter. It is fully dependent on $\vec c$ and the filling factor, and can in principle be obtained in numerics by finite size scaling. 

\subsection{$k_0=5$}
We now look at model Hamiltonians with pseudopotential combinations of $\hat V_1^{\text{2bdy}}, \hat V_3^{\text{2bdy}}$ and $\hat V_5^{\text{2bdy}}$. For simplicity we will only look at the case of $c_1=c_1^{\lambda_1},c_3=c_3^{\lambda_1},c_5=c_5^{\lambda_1}$. While this does not cover all possible cases, it gives much insight into the behaviours of such model Hamiltonians. Similar to the case of $k_0=3$, we can obtain the following relationship:
\begin{eqnarray}\label{k05}
&&\delta E_{\bm q}\propto \lambda_1 E_0+42\left(c_5d_7-c^{\lambda_1}_7d_5\right)\nonumber\\
&&+\left(\lambda_2 E_0-546\left(c_5^{\lambda_2}d_7-c_7^{\lambda_2}d_5\right)\right)q^2\nonumber\\
&&+\left(c_5^{\lambda_1}-c_5^{\lambda_2}\right)\left(682d_5-180d_3-546d_7\right)q^2+O(q^4)\qquad\quad
\end{eqnarray}
Here $\lambda_1,\lambda_2,c_5^{\lambda_1},c_5^{\lambda_2}$ are defined the same way as in Eq.(\ref{k03}), while $c_7^{\lambda_1}=-2/5+57c_3/70+19c_3^2/35+3c_3^3/70, c_7^{\lambda_2}=-166/819+18899c_3/24570+23c_3^2/91+5c_3^3/546$. Comparing to Eq.(\ref{k03}) we have an additional parameter $d_3$.
\begin{figure}
\includegraphics[width=\linewidth]{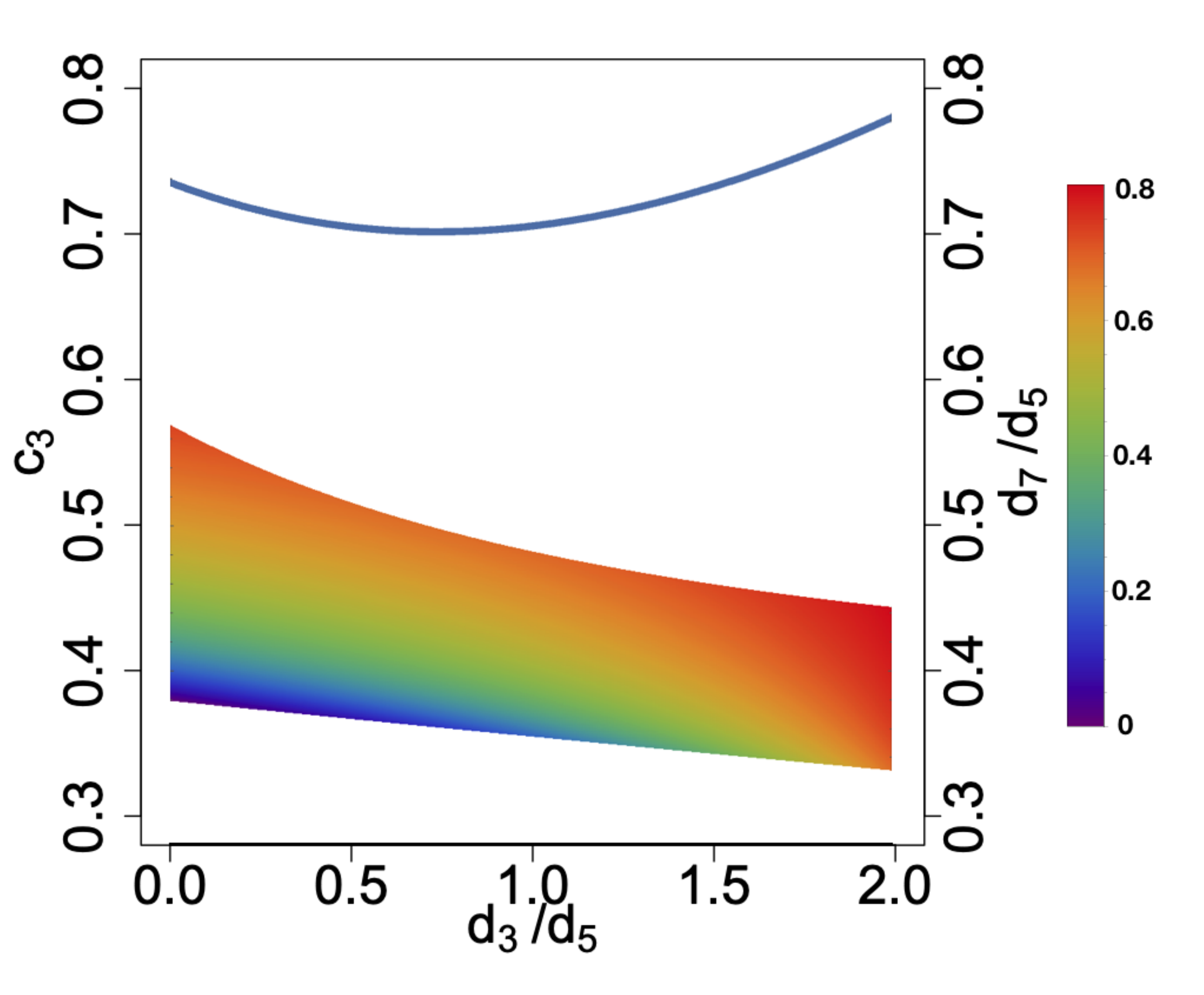}
\caption{The range of parameters where the FQHN is possible at different values of $d_3/d_5$, as given by Eq.(\ref{k05}). The shaded area is the range of $c_3$ as given by the left axis. The line plot is the maximum allowed value of $d_7/d_5$ at different value of $d_3/d_5$, as given by the right axis. The heat map gives the maximum allowed value of $d_7/d_5$ for different values of $c_3$ and $d_3/d_5$. }
\label{fig3}
\end{figure} 
For different values of $d_3/d_5$, we can determine the respective narrow ranges of parameter space for $c_3$ and $d_7/d_5$, in which the FQHN phase is possible. This is captured in Fig.(\ref{fig3}), for the range of $0<d_3/d_5<2$. Not much can be done analytically at this stage, though at any specific filling factor, some numerical computations can be performed to explore the possibility of the FQHN phases at different values of $\lambda_1$. 

\section{Numerical Studies}\label{ms}

All results in Sec.~\ref{mm} are valid in the thermodynamic limit, and are applicable at any filling factor. In this section, we perform some preliminary numerical analysis at filling factor $\nu=1/3$, about possible microscopic models for the FQHN. While we are looking at a particular filling factor, the methodologies for the numerical analysis described here can be applied to any filling factors. A more specialised analysis of the FQHN states for the Laughlin phase will be given in the next section. 

We will show that the analytic derivations from the previous sections can strongly constrain the parameter space for the realisation of FQHN in the thermodynamic limit. Thus numerically, we only need to look for the FQHN phase over a much smaller parameter space in the form of the linear combination of pseudopotentials. All numerical computations in this work are done with the spherical geometry\cite{haldane4}, and we analyse the ground state wavefunctions and energy spectra for reasonably large system sizes. While the comparison between finite size scaling of numerical results and the analytical results in Sec.~\ref{mm} can never be conclusive, the results here nevertheless illustrate the usefulness and limitations of finite size numerical calculations.

The neutral gap of $\delta E_{\bf q\rightarrow 0}$ in this section is not computed from the energy spectrum. Instead we use Eq.(\ref{compact1}) to evaluate the energy gap numerically from the ground state in the $L=0$ sector alone. Not only is this a simpler calculation technically, it also has smaller finite size effect. This is because $\Gamma_{\left(1\right)}$ is calculated from the thermodynamic limit and the only size dependent quantity is $\bm d$. In addition, it allows us to compute the energy gap in the limit $\bf q\rightarrow 0$, which is inaccessible from the full spectra of the finite systems.

\begin{figure}
\includegraphics[width=\linewidth]{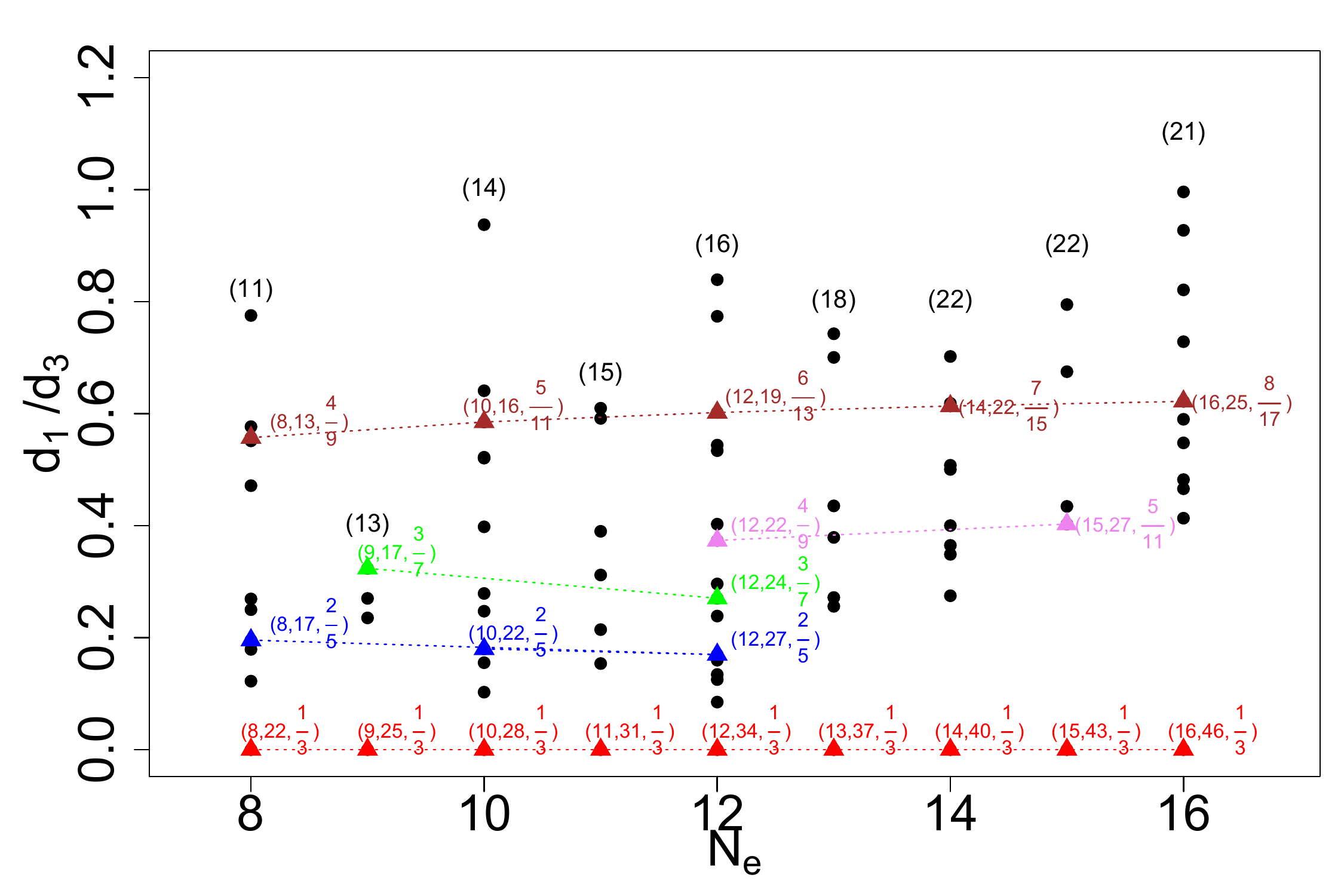}
\caption{The value of $d_1/d_3$, computed from the ground state of the $\hat V_1^{\text{2bdy}}$ interaction at different Hilbert spaces (indexed by the number of electrons $N_e$ and number of orbitals $N_o$.). The Jain series are highlighted with different colors, where the numbers in the brackets are $\left(N_e, N_o,\nu\right)$. The number on top of each $N_e$ sector is the minimum number of $N_o$ included in the plot; for smaller $N_o$ not included in this plot we have $d_1/d_3>1$.}
\label{fig1}
\end{figure} 

For the model Hamiltonian consisting of only $\hat V_1^{\text{2bdy}}$ pseudopotential (i.e. $c_1=1, c_{i>1}=0$, or the $k_0=1$ model), there are no tuning parameters, and the variational energy gap of the SMA state is completely controlled by $d_3-d_1$ (see Eq.(\ref{v1})). In Fig.(\ref{fig1}), we scan over all possible combinations of $N_e, N_o$ that are numerically accessible, and compute $d_1, d_3$ from the ground state in the $L=0$ sector (not necessarily the global ground state). The numerical results show strong evidence that for any FQH phases that can potentially be supported by the $k_0=1$ model (which in particular includes many Abelian Jain states), the SMA states in the long wavelength limit is gapped from the ground state in the $L=0$ sector.

An interesting observation is that with the $\hat V_1^{\text{2bdy}}$ model Hamiltonian and for all finite size systems we have accessed, the global ground state is in the $L=0$ sector \emph{if and only if} the filling factor and the topological shift corresponds to the Jain series, i.e. $N_o=\left(2n+1\right)N_e/n-n-1$, and their particle-hole conjugates. These Hilbert spaces are highlighted in Fig.(\ref{fig1}). For all of these cases we have $d_1/d_3<1$, indicating gapped neutral excitations as $|q|\rightarrow 0$. For other combinations of $N_e, N_o$ where the global ground state is not in the $L=0$ sector, it could be because the neutral excitations go soft even for finite size systems (probably of unknown filling factors), and they could still have a charge gap. However in all cases where $N_o$ is reasonably large, $d_1/d_3<1$ as well. There is thus no numerical evidence of the FQHN. For each $N_e$, $d_1/d_3>1$ only when $N_o$ is rather small. This implies as $N_e$ increases, we can only have $d_1/d_3>1$ at rather large filling factors ($\nu\gtrsim 0.75$).

\begin{figure}
\includegraphics[width=\linewidth]{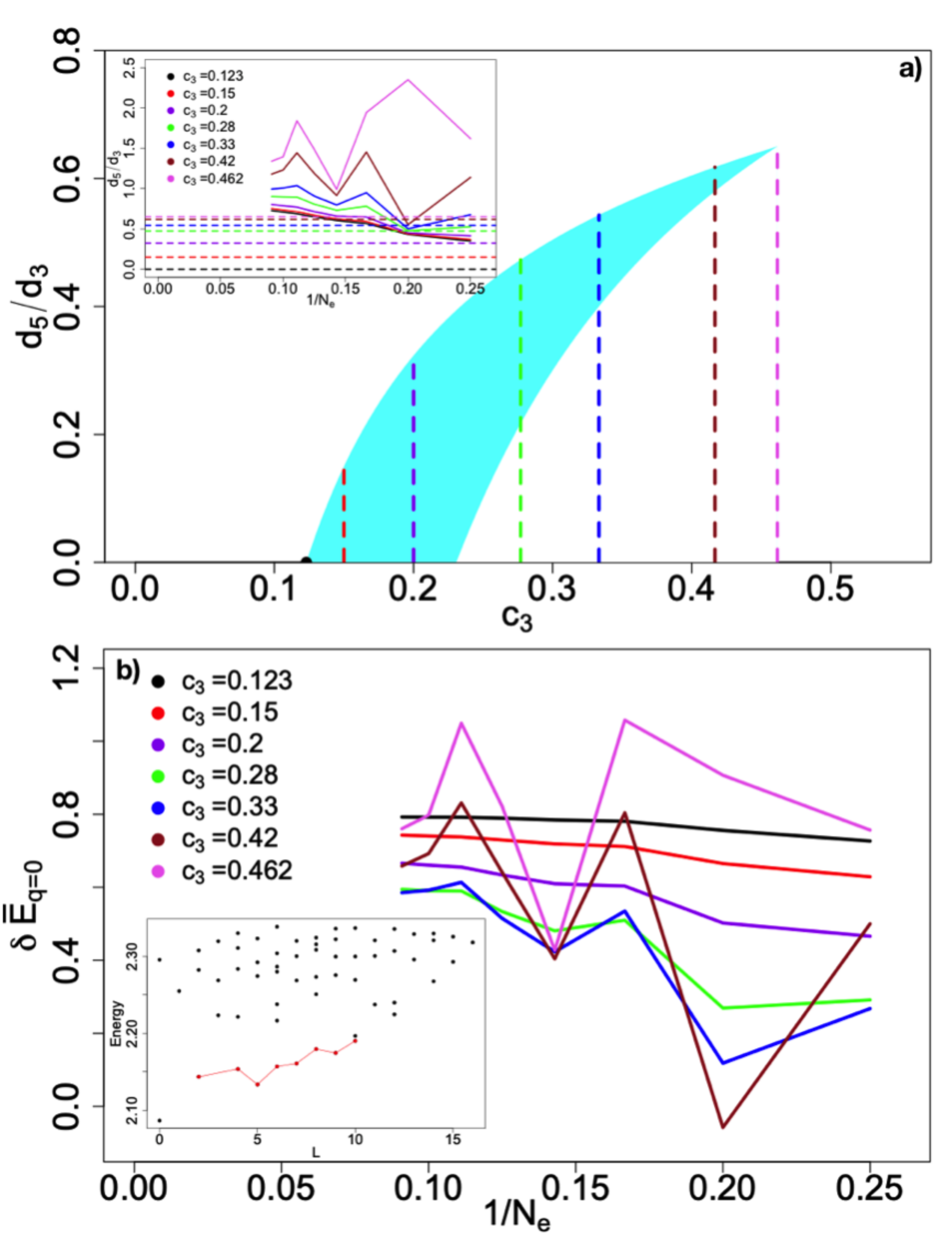}
\caption{a). The shaded region is the possible values of $\left(c_3,d_5/d_3\right)$ for the FQHN phase, based on the analytical results of Eq.(\ref{k03}) in the thermodynamic limit. The upper end of the vertical dotted lines gives the upper bound of $d_5/d_3$ at different values of $c_3$, also given by the horizontal lines in the inset with the same color code. The inset also shows the scaling of $d_5/d_3$ for different system sizes at different values of $c_3$. b). The scaling of Eq.(\ref{k03}) for different system sizes and different values of $c_3$. The inset is the energy spectrum for $c_3=0.462$, and the low-lying neutral modes are highlighted in red.}
\label{fig2}
\end{figure}

We now move onto $k_0=3$ model Hamiltonians that are linear combinations of $\hat V_1^{\text{2bdy}}$ and $\hat V_3^{\text{2bdy}}$ pseudopotentials, where without loss of generality we set $c_1=1$. From the general expression of Eq.(\ref{k03}), the allowed range of $c_3$ and $d_5/d_3$ is given in the shaded area in Fig.(\ref{fig2}a), which is computed analytically in the thermodynamic limit. In particular, the FQHN phase is not possible in the thermodynamic limit for $0.123<c_3<0.462$. Any numerical evidence suggesting otherwise is due to finite size effects. 

For the Laughlin phase at $\nu=1/3$, finite size analysis is carried out at different values of $c_3$, at which $d_5/d_3$ is computed from the ground state (in the $L=0$ sector). They show that it is very unlikely for $d_5/d_3$ to be below the maximally allowed value (see Fig.(\ref{fig2}a) inset) at all possible values of $c_3$. The scaling shows that the variational energy gap (first line of Eq.(\ref{k03})) also seems to be finite, which is consistent (see Fig.(\ref{fig2}b)). While the finite size energy spectrum does seem to indicate softening of the neutral mode in the long wavelength limit (see Fig.(\ref{fig2}b) inset) for some values of $\lambda_1$, that most likely will not be the case when larger system sizes become accessible numerically. 

For $k_0=5$ model Hamiltonians, the addition of $\hat V_5^{\text{2bdy}}$ introduces additional variables $c_5$ and $d_7$, making thorough numerical investigation difficult. We look at the special case when $\vec c$ comes from the eigenvectors of $\Gamma_{\left(1\right)}$, i.e. $c_i=c_i^{\lambda_1}$ for $i=1,3,5$, and $c_{i>5}=0$. From the analytic expression of Eq.(\ref{k05}), we have rigorous results on the range of $c_3$ in different scenarios. At filling factor $\nu=1/3$ numerical computations show it is unlikely for $d_3/d_5>2$ in the thermodynamic limit for a wide range of $c_1>c_3>c_5$. For each value of $d_3/d_5$, we can analytically calculate the possible range of $c_3, d_7/d_5$ from Eq.(\ref{k05}) for the FQHN phase. The results are plotted in Fig.(\ref{fig3}). In particular, only a small range of $c_3$ needs to be explored for the potential realisation of the FQHN at $\nu=1/3$.

\begin{figure}
\includegraphics[width=\linewidth]{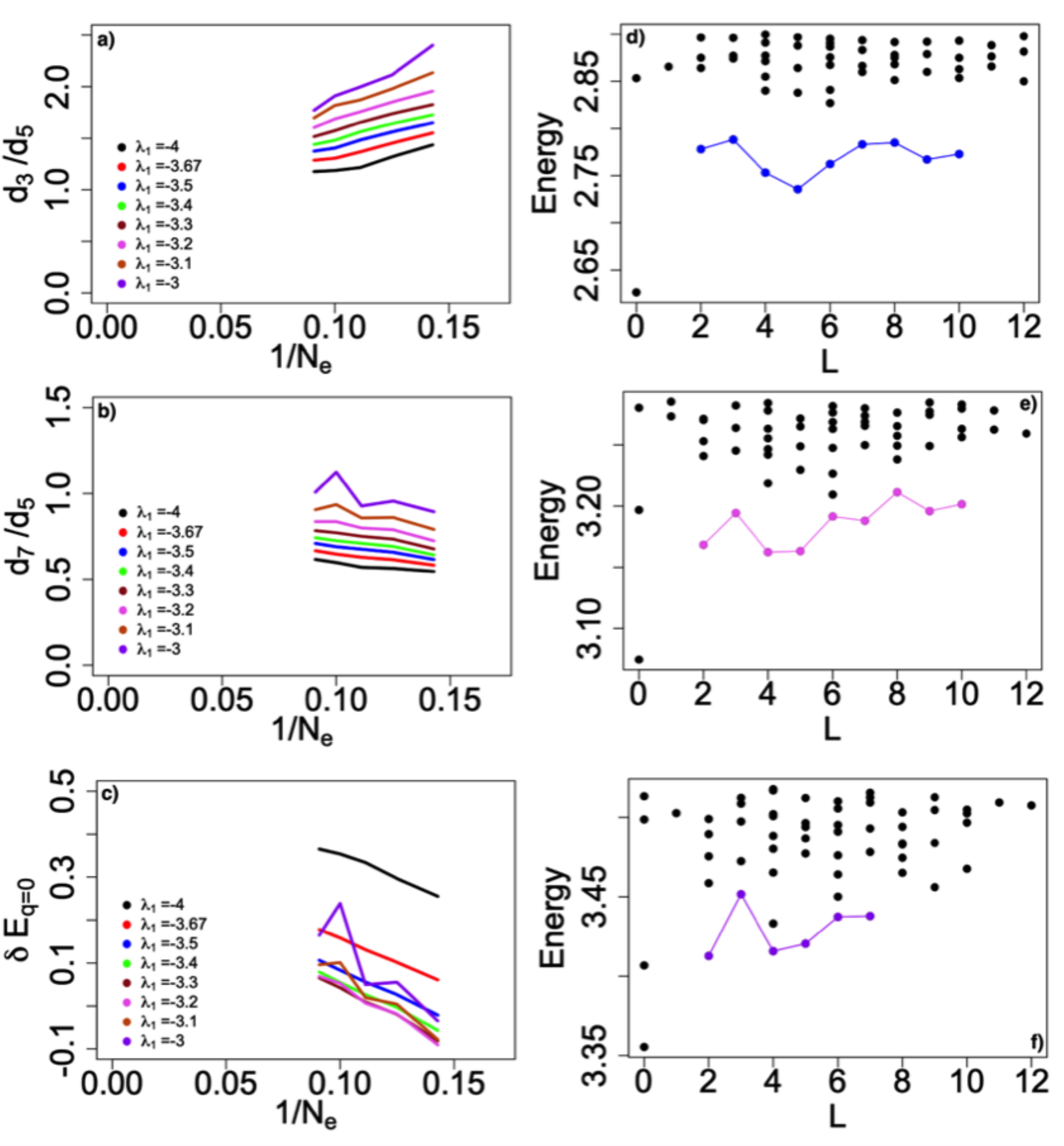}
\caption{The left panels are finite size scaling of a). $d_3/d_5$; b). $d_7/d_5$; c). numerator of $\delta E_{\bm q=0}$ for the $k_0=5$ models with different values of $\lambda_1$, corresponding to different set of values of $c_1, c_3, c_5$. The right panels give the energy spectra with $11$ electrons and $31$ orbitals for d). $\lambda_1=-3.5$; e). $\lambda_1=-3.2$; f). $\lambda_1=-3$. The low-lying neutral excitations are highlighted.}
\label{fig4}
\end{figure}

Since there is a unique relationship between $\lambda_1$ and $\left(c_3,c_5\right)$, different values of $\lambda_1$ are plotted in Fig.(\ref{fig4}), by diagonalising the full Hilbert space and extracting $d_i$ from the corresponding ground states. In the limit of $N_e\rightarrow\infty$, $d_3/d_5$ seems to fall in between $0.5$ and $1.5$ (see Fig.(\ref{fig4}a)). From Fig.(\ref{fig3}) we thus need $d_7\sim 0.7$ (without being too precise), and this also seems quite possible from Fig.(\ref{fig4}b). For $\lambda_1<-3.5$, the finite size scaling of Eq.(\ref{k05}) seems to clearly indicate that $\delta E_{\bm q\rightarrow 0}$ does not go soft. On the other hand, for $\lambda_1>-3.2$, the finite size effect becomes strong, potentially indicating the divergence of the ground state correlation length and the closing of the neutral gap in the long wavelength limit (see Fig.(\ref{fig4}c)). 

We thus expect the minimal microscopic model for the FQHN at $\nu=1/3$ consists of a linear combination of $\hat V_1^{\text{2bdy}}, \hat V_3^{\text{2bdy}}, \hat V_5^{\text{2bdy}}$. The results here apply to zero temperature, where spontaneous symmetry breaking can only happen at $\delta E_{\bm q\rightarrow 0}=0$. It is possible to have a range of parameters for the FQHN phase to be stable, especially if the interaction is allowed to be more long ranged. At finite temperature, FQHN phase can be observed as long as the neutral excitation gap $\delta E_{\bm q\rightarrow 0}$ is much smaller than the charge gap $\Delta E_c$, and the former is smaller than the thermal energy $k_BT$, i.e. $\delta E_{\bm q\rightarrow 0}\ll k_BT\ll\Delta E_c$. Thus in realistic experimental setting, the FQHN phase could be stable against disorder and small perturbations, as long as the charge gap is the dominant energy scale.

\section{nematic fractional quantum hall for the Laughlin phase}\label{qd}
While the previous sections describe analytic and numerical methodologies for studying generic FQHN phases at any filling factors, in this section we reveal more special properties of the Laughlin phase at $\nu=1/3$ that are relevant to the FQHN. Not only does these special properties allow us to extend the general results derived in the previous sections, it also leads to much better understandings of the nature of the softening of the neutral modes at $\nu=1/3$. A number of favourable experimental conditions are also proposed, which can lead to more robust realisation of the FQHN phase and even the observation of Haffnian-like FQH states\cite{hf1,hf2}, as we will explain below.

\subsection{Elementary excitations of the Laughlin phase}
It is instructive to first go over the elementary neutral excitations for the Laughlin phase. The low-lying neutral excitations of the Laughlin phase have been well-studied\cite{yangbo1,yangbo2,sreejith,ajit1,ajit2,yangbo3,haldane4}. In the long wavelength limit, the neutral excitations are quadrupole excitations well approximated by the projected density mode, or the single mode approximation\cite{gmp}. The model wavefunctions of the entire branch of the neutral excitations (also called the magnetoroton mode, with quadrupole excitations at small momenta and dipole excitations at large momenta) can be constructed either using the Jack polynomial formalism\cite{yangbo1} and the corresponding first quantised wavefunctions\cite{yangbo2}, or using exciton states in the composite fermion picture\cite{sreejith,ajit1,ajit2}. At large momenta, a magnetoroton mode is a neutral excitation that consists of a pair of well separated quasielectron and quasihole. The separation increases with the momenta, together with its dipole moment. The interaction between the quasielectron and quasihole thus becomes negligible at large momenta, and the neutral excitation energy is equal to the energy of creating one quasielectron and one quasihole (each in isolation). Thus the charge gap is more or less the same as the neutral dipole excitations at large momenta.

As the momentum of the magnetoroton mode decreases, so is the separation between the quasielectron and quasihole (and thus its dipole moment). In the long wavelength limit, the dipole moment vanishes, with the quasielectron and quasihole merging to form a uniform geometric deformation of the Laughlin ground state. Such excitations do have non-vanishing quadrupole moments. These characteristics of the neutral excitations are universal and not just limited to the Laughlin phase. 

While the magnetoroton mode forms a continuous band of dispersion, it is clear the excitation at the long wavelength limit is qualitatively different from that at large momenta. This is in particular reflected in their dynamical properties as we will show shortly. The physical intuitions on how the dynamics of the quadrupole and dipole excitations can be affected by microscopic interaction can be made more transparent with the root configurations of their model wavefunctions as follows\cite{bernevig,yangbo1}:
\begin{eqnarray}\label{root1}
&&\d{1}\d{1}0\textsubring{0}\textsubring{0}0100100100100\cdots\qquad\text{L=2} \nonumber\\
&&\d{1}\d{1}0\textsubring{0}010\textsubring{0}0100100100\cdots\qquad\text{L=3} \nonumber\\
&&\d{1}\d{1}0\textsubring{0}010010\textsubring{0}0100100\cdots\qquad\text{L=4} \nonumber\\
&&\d{1}\d{1}0\textsubring{0}010010010010\textsubring{0}0\cdots\qquad\text{L=5} 
\end{eqnarray}
Here each root configuration represents a monomial, or a Slater determinant given by the occupation of orbitals in a single LL. The digits going from left to right correspond to orbitals going from the north pole to the south pole on the spherical geometry, and ``1" means the orbital is occupied, while ``0" means the orbital is un-occupied by the electron. The solid and open circles beneath the digits indicate the locations of quasiparticles (of charge $e/3$, when three consecutive orbitals contain more than one electron) and quasiholes (of charge $-e/3$, when three consecutive orbitals contain fewer than one electron). Each root configuration represents a many-body wavefunction, where only basis ``squeezed" from the root configuration have non-zero coefficients. The $L$ sector to the right of the root configuration is the total angular momentum quantum number on the sphere. The quadrupole excitation is given by the state with $L=2$, while the dipole excitations are given by $L>0$.

The root configurations clearly show the increasing separation of the quasielectron (clustered to the left, or the north pole) from the quasihole, as the angular momentum increases. They also encode dynamical properties of the excitations, as we will show below.

\subsection{The Haffnian and the quadrupole excitation}
The model wavefunction for the quadrupole excitation (with $L=2$) is the exact zero energy state of the Haffnian model Hamiltonian (consisting of the linear combination of three-body pseudopotentials $\hat V_3^{\text{3bdy}},\hat V_5^{\text{3bdy}},\hat V_6^{\text{3bdy}}$). In contrast, the model wavefunctions for the dipole excitations (with $L>2$) are the exact zero energy state of the Gaffnian\cite{simon} model Hamiltonian (consisting of the linear combination of three-body pseudopotentials $\hat V_3^{\text{3bdy}},\hat V_5^{\text{3bdy}}$). These did not seem to be recognised before in the literature, but are easy to see from the recently developed LEC formalism\cite{lec} (i.e. the $L=2$ state satisfies the LEC condition $\{2,1,2\}\lor\{6,2,6\}$, while the $L>2$ state satisfies the LEC condition $\{2,1,2\}\lor\{5,2,5\}$), using the root configurations in Eq.(\ref{root1}) and the associated squeezed basis.

Using this insight, one can consider a theoretical model with the following Hamiltonian:
\begin{eqnarray}\label{v1vh}
\hat{\mathcal H}_\lambda=\left(1-\lambda\right)\hat V_1^{\text{2bdy}}+\lambda\hat{\mathcal H}_{\text{haff}}
\end{eqnarray}
where $\hat V_1^{\text{2bdy}}$ is the model Hamiltonian for the Laughlin state at $\nu=1/3$ in the form of the Haldane pseudopotential, while $\hat{\mathcal H}_{\text{haff}}=\hat V_3^{\text{3bdy}}+h_5\hat V_5^{\text{3bdy}}+h_6\hat V_6^{\text{3bdy}}$ is the Haffnian model Hamiltonian with $h_5,h_6>0$. 

Since $\hat{\mathcal H}_{\text{haff}}$ gives an energy punishment for all of the $L>2$ neutral excitations, but not for the quadrupole excitation at $L=2$, we expect the softening of the quadrupole excitation as $\lambda$ increases. At $\lambda=1$ the quadrupole excitation will be exactly degenerate with the Laughlin state (both with zero energy) even for finite systems. Eq.(\ref{v1vh}) can thus be considered as the model Hamiltonian capturing the essential physics for the transition from the fully gapped Laughlin phase (at $\lambda=0$) to the FQHN phase when the neutral mode goes soft (at $0<\lambda<1$, since in the thermodynamic limit the gap may close at some intermediate value of $\lambda$). We can now see that the FQHN phase at $\nu=1/3$ is related to the Haffnian phase, which also occurs at $\nu=1/3$ but with a different topological shift as compared to the Laughlin phase. While $\hat{\mathcal H}_{\text{haff}}$ is conjectured to be gapless from the conformal field theory perspective, a finite gap may open as $\lambda$ decreases from $1$ (in analogy to the gap opening away from the Gaffnian model Hamiltonian\cite{jolicoeur}), leading to an incompressible ground state with different topological properties (though the quasihole excitations may not be non-Abelian\cite{yangbo4}). This interesting connection will be explored in the future works.

\begin{figure}
\includegraphics[width=\linewidth]{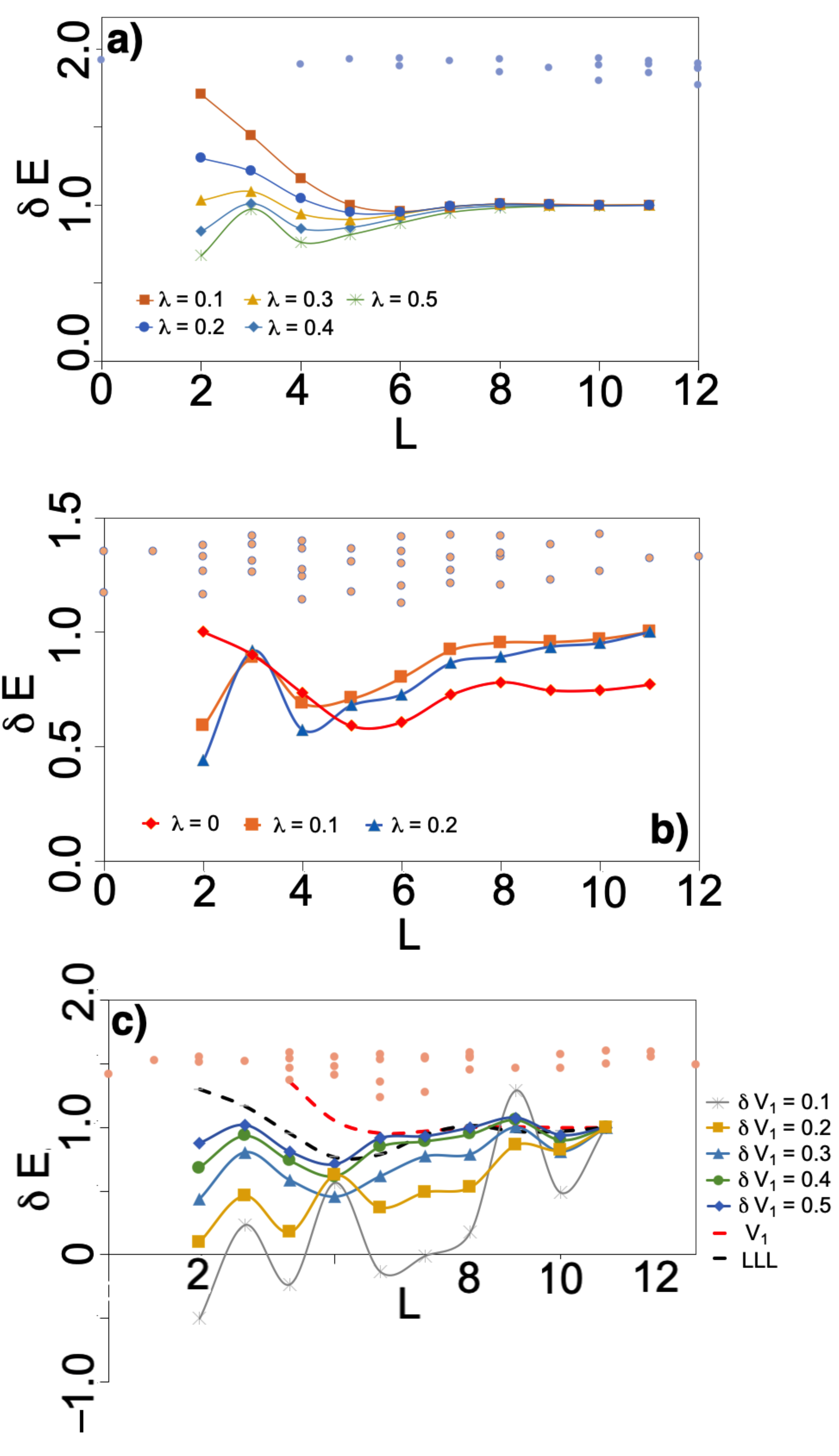}
\caption{The energy spectrum of various model Hamiltonians from exact diagonalisation with $11$ electrons and $31$ orbitals, with $h_5=h_6=1$. The ground states in the $L=0$ sector are set to zero, and the ground states in the $L=11$ sector are normalised to unity. a). The spectra of Eq.(\ref{v1vh}). b). The spectra of Eq.(\ref{vlllvh}). c). The spectra of Eq.(\ref{vh2bdy}), modified by a small $\delta V_1$. Only the multi-roton continuum from $\delta V_1=0.5$ is included in the plot to avoid clutters}
\label{fig5}
\end{figure}

In Fig.(\ref{fig5}a) we look at the special case of $h_5=h_6=1$ (other positive values give qualitatively same results). As we tune $\lambda$ away from zero in Eq.(\ref{v1vh}), there is very clean numerical evidence of the quadrupole excitations going soft, while the entire magnetoroton mode branch is well separated from multi-roton continuum for all the spectra in the figure. All energies are measured from the ground state energy in the $L=0$ sector. The ground state energies in the $L=11$ sector (again measured from the $L=0$ ground state energies) are normalised to unity. This is the sector where a single quasielectron-quasihole pair is maximally separated for the system size of $11$ electrons, so its energy can be considered as the charge/dipole excitation gap. Thus Eq.(\ref{v1vh}) shows that the neutral gap can be much smaller than the charge gap even for finite systems.

\subsection{The experimental relevance}
It is interesting to look more realistic interactions, using the following model:
\begin{eqnarray}\label{vlllvh}
\hat{\mathcal H}_\lambda=\hat V_{\text{LLL}}+\lambda\hat{\mathcal H}_{\text{haff}}
\end{eqnarray}
where $\hat V_{\text{LLL}}$ is the lowest Landau level two-body Coulomb interaction. In this case, a very small amount of three-body interaction (which can come from LL mixing) will significantly soften the quadrupole excitations, as one can see from Fig.(\ref{fig5}b). We suspect the similar is also true in the second Landau level, but the numerical spectrum tends to be too messy due to the strong finite size effect (given the more long-range interaction). The results here do suggest that LL mixing can play a very significant role for the FQHN in realistic systems.

The new understandings of the quadrupole excitations at $\nu=1/3$ allows us to propose realistic two-body interactions that favour the FQHN phase in various experimental settings. At large magnetic field when Landau level (LL) mixing is suppressed, we only have effective two-body interactions. A useful two-body interaction can be proposed as follows:
\begin{eqnarray}\label{vh2bdy}
\hat V_{\text{haff}}^{\text{2bdy}}=\hat{\mathcal H}_{\text{haff}}+\hat{\mathcal H}^a_{\text{haff}}
\end{eqnarray}
where $\hat{\mathcal H}^a_{\text{haff}}$ is the particle-hole conjugate of $\hat{\mathcal H}_{\text{haff}}$, and in the thermodynamic limit we have
\begin{eqnarray}
\hat V_{\text{haff}}^{\text{2bdy}}=&&\hat V_1^{\text{2bdy}}+\frac{2}{3}\left(\frac{2+h_5+5h_6}{4+3h_5+h_6}\right)\hat V_3^{\text{2bdy}}\nonumber\\
&&+\left(1-\frac{4}{3}\left(\frac{3+h_5}{4+3h_5+h_6}\right)\right)\hat V_5^{\text{2bdy}}
\end{eqnarray}
where we normalise the coefficient of $\hat V_1^{\text{2bdy}}$ to be unity. This family of two-body interactions are expected to retain most of the correlation properties of the Haffnian model Hamiltonian, and indeed at the topological shift of the Laughlin phase ($N_o=3N_e-2$), the ground states in the $L=0$ and $L=2$ sectors are very close in energy. For large systems the global ground state tends to be in the $L=2$ sector We can thus expect a phase transition when the $L=0$ and $L=2$ sectors become degenerate, by small modifications of $\hat V_1^{\text{2bdy}}$ in $\hat V_{\text{haff}}^{\text{2bdy}}$. In Fig.(\ref{fig5}c), we add a small amount of $\hat V_1^{\text{2bdy}}$ to Eq.(\ref{vh2bdy}) to monitor the dispersion of the magnetoroton mode. While we do not have a well-separated gap between the magnetoroton mode and the multi-roton continuum for the finite systems, the behavours of the magnetoroton mode are qualitatively similar to Fig.(\ref{fig5}a,b).

The resulting two-body interaction for stabilising the FQHN phase agrees qualitatively with what we obtained from the general approach in Sec.\ref{ms}. It also gives a better understanding why the minimal models for the FQHN at $\nu=1/3$ should consist of $\hat V_1^{\text{2bdy}}, \hat V_3^{\text{2bdy}}, \hat V_5^{\text{2bdy}}$, since it is derived from the parent Hamiltonian of the Haffnian state with two freely tunable parameters $h_5,h_6$. Given that with the $\hat V_1^{\text{2bdy}}$ and LLL Coulomb interaction the quadrupole excitation energy is high up in the continuum (see the black and red dotted line plots in Fig.(\ref{fig5}b)), the FQHN phase requires interaction to be more long range than $\hat V_1^{\text{2bdy}}$, but shorter range than LLL Coulomb. This can be achieved at large magnetic field by properly tuning the sample thickness and/or dielectric screening. With the model Hamiltonians of the FQHN phase, the desirable range of experimental parameters can now be computed analytically.

In higher LLs, the two-body interaction is long ranged and definitely differs more from the model FQHN Hamiltonian. While this does not preclude the realisation of the FQHN phase, numerical analysis becomes more difficult due to stronger finite size effect. However, we also expect stronger LL mixing at higher LLs. The connection to the Haffnian model Hamiltonian clearly suggests that LL mixing could be helpful in realising the FQHN phase. Note that $\hat V_3^{\text{3bdy}}, \hat V_5^{\text{3bdy}}$ do not affect the quadrupole or dipole excitations, since they all live in the null space of these two pseudopotentials. It is $\hat V_6^{\text{3bdy}}$ that is playing the important role here, since it is the only pseudopotential that punishes the dipole excitations, but not the quadrupole excitations. Starting from the fully gapped Laughlin phase, we thus need a small positive $\hat V_6^{\text{3bdy}}$ to push it into the FQHN phase with a vanishing neutral gap.

The effective two-body and three-body interactions in the pseudopotential basis can be analytically computed for realistic samples with various tuning parameters (e.g. sample thickness, screening, band dispersion, in-plane magnetic field, etc.)\cite{yangbo5}. One can design suitable samples for the robust realisation of the FQHN based on detailed calculations, which we will present elsewhere. In general, we would like the two-body interaction to be dominated by $\hat V_1^{\text{2bdy}}$, but with vanishing long range part from $\hat V_n^{\text{2bdy}}$ with $n>5$. When the LL mixing effect is not negligible (e.g. intermediate magnetic field), we would like a positive $\hat V_6^{\text{3bdy}}$ to further stabilise the FQHN. A negative $\hat V_6^{\text{3bdy}}$, on the other hand, could open the neutral gap at $\nu=n+1/3$. However, it could still favour FQHN at $\nu=n+2/3$, which is where the anti-Laughlin phase is realised.

\section{Conclusions}\label{con}

We have computed analytically the dynamical behaviours of the neutral excitations in the long wavelength and thermodynamic limit, which is applicable to any FQH phase with a charge gap. Such excitations are quadrupole excitations, with its gap and dispersion relations captured by two universal tridiagonal matrices that are independent of microscopic details. Both the nematic order and the gapless modes from spatially varying nematic order are studied, and we can show that such nematic wave dispersion is quadratic at the quantum critical point. Specific criteria for the FQHN phase to be robust are also derived, which are necessary (though not sufficient) conditions for the coexistence of the anisotropic transport and the topologically protected Hall conductivity plateau.

In addition, we show that the gap of the quadrupole excitation and its dispersion in long wavelength limit can be completely determined from the ground state properties, given the universality of the characteristic matrices in the thermodynamic limit. This provides a new approach in studying the potential transition from isotropic to FQHN phases both analytically and numerically at the microscopic level. Numerical analysis on the Laughlin phase at filling factor $\nu=1/3$ for reasonably large system sizes shows evidence that the phase transition is only likely for microscopic Hamiltonians that are linear combinations of at least \emph{three} leading Haldane pseudopotentials (i.e. $\hat V_1^{\text{2bdy}}, \hat V_3^{\text{2bdy}}, \hat V_5^{\text{2bdy}}$). We also show at this filling factor the FQHN is strongly connected to the Haffnian phase, a competing topological phase at the same filling factor but with a different topological shift. The analytical results can narrow down the parameter range for the short range interactions, allowing us to see tentative evidence of the softening of the neutral excitations in the fermionic systems. Several favourable experimental conditions are proposed where the neutral gap in the long wavelength limit is likely to be much smaller than the charge gap.

The characteristic matrices derived in this work shows that the dynamics of the quadrupole excitations has universal aspects that can potentially be useful for constructing effective theories for the FQH effects. From the microscopic perspective, more work is needed to fully understand the competition between the quadrupole gap and dipole gap for different FQH phases. The latter essentially gives the charge gap of the FQH fluid, and needs to be finite for the quantum fluid to be incompressible. Our results tentatively suggests that while short range interaction generally support a finite charge or dipole gap, it can nevertheless lead to softening of the quadrupole gap. For example the $k_0=5$ model Hamiltonians we analysed in the paper has a much shorter range than the lowest Landau level Coulomb interaction. On the other hand, for very short range interaction (e.g. $k_0=1$ or $k_0=3$ models), the quadrupole gap becomes very large and merge into the multi-roton continuum. The underlying physics of such behaviours is still not well understood.

In our numerical analysis we ignored the $q^4$ coefficient of the structure factor, which is the denominator of the quadrupole gap. This should be justified since it is bounded from below by the Hall viscosity\cite{haldane3}. Perturbation from the $V_1$ model Hamiltonian should only have the possibility of increasing the coefficient (thus reducing the quadrupole gap further). Nevertheless, more detailed numerical analysis is needed to see if including the $q^4$ coefficient can give clearer finite scaling of various aspects of the quadrupole excitations. The results in this work is also applicable for any filling factors. It is interesting to explore the possibility of the FQHN phases in other filling factors, especially for the non-Abelian phases where there are multiple branches of the low-lying neutral modes.

\begin{acknowledgments}
I thank Zlatko Papic for pointing my attention to the nematic Goldstone modes, Ajit Balram, Xin Wan and Zhao Liu for useful discussions, as well as the referees for the constructive comments. This work is supported by the Singapore National Research Foundation (NRF) under NRF fellowship award NRF-NRFF12-2020-0005.
\end{acknowledgments}

\clearpage
\begin{widetext}
\begin{center}
\textbf{\large Supplementary Online Materials for ``Microscopic theory for the nematic fractional quantum Hall effect"}
\end{center}
\setcounter{equation}{0}
\setcounter{section}{0}
\setcounter{figure}{0}
\setcounter{table}{0}
\setcounter{page}{1}
\makeatletter
\renewcommand{\theequation}{S\arabic{equation}}
\renewcommand{\thefigure}{S\arabic{figure}}
\renewcommand{\bibnumfmt}[1]{[S#1]}
\renewcommand{\citenumfont}[1]{S#1}

In this supplementary material, we give more technical details for the analytical computations in the main text. 

\section{Long wavelength variational energy of neutral excitations}
Starting with SMA trial wavefunction $|\psi_{\bm q}\rangle=\delta\hat\rho_{\bm q}|\psi_0\rangle$, where $\delta\hat\rho$ is defined in the main text and $|\psi_0\rangle$ is the ground state of the two-body effective Hamiltonian $\mathcal H$ in the $\bm q=0$ sector (where $\bm q$ is the linear momentum), we compute the following commutation:
\begin{eqnarray}
[\delta\hat\rho_{-\bm q},[\hat{\mathcal H},\delta\hat\rho_{\bm q}]]&=&\int\frac{d^2q'}{4\pi}V_{\bm q'}[\delta\hat\rho_{-\bm q},[\hat\rho_{\bm q'}\hat\rho_{-\bm q'},\delta\hat\rho_{\bm q}]]\nonumber\\
&=&\int\frac{d^2q'}{4\pi}V_{\bm q'}\left([\delta\hat\rho_{-\bm q},[\delta\hat\rho_{\bm q'}\delta\hat\rho_{-\bm q'},\delta\hat\rho_{\bm q}]]+2\pi\delta^2\left(\bm q'\right)[\delta\hat\rho_{-\bm q},[\delta\hat\rho_{-\bm q'},\delta\hat\rho_{\bm q}]]\right)
\end{eqnarray}
Using the GMP algebra, the second term vanishes, we thus have:
\begin{eqnarray}
\langle\psi_0[\delta\hat\rho_{-\bm q},[\hat{\mathcal H},\delta\hat\rho_{\bm q}]]|\psi_0\rangle&=&2\langle\psi_{\bm q}|\hat{\mathcal H}|\psi_{-\bm q}\rangle-2E_0S_{\bm q}\\
&=&\int\frac{d^2q'}{4\pi}V_{\bm q'}\langle\psi_0|[\delta\hat\rho_{-\bm q},[\delta\hat\rho_{\bm q'}\delta\hat\rho_{-\bm q'},\delta\hat\rho_{\bm q}]]|\psi_0\rangle\nonumber\\
&=&\int\frac{d^2q'}{4\pi}V_{\bm q'}\left(2\sin\left(\frac{1}{2}\bm q'\times\bm q\right)\right)^2\left(s_{\bm q'+\bm q}+s_{\bm q'-\bm q}-2s_{\bm q'}\right)
\end{eqnarray}
Here $E_0$ is the ground state energy given by $\hat{\mathcal H}|\psi_0\rangle=E_0|\psi_0\rangle$, and $s_{\bm q}=S_{\bm q}-S_\infty$ as defined in the main text. Thus the variational energy of $|\psi_{\bm q}\rangle$ gap is given by:
\begin{eqnarray}
\delta E_{\bm q}&=&E_{\bm q}-E_0=\frac{\langle\psi_{\bm q}|\hat{\mathcal H}|\psi_{\bm q}\rangle}{\langle\psi_{\bm q}|\psi_{\bm q}\rangle}-E_0=\frac{1}{2S_{\bm q}}\langle\psi_0[\delta\hat\rho_{-\bm q},[\hat{\mathcal H},\delta\hat\rho_{\bm q}]]|\psi_0\rangle\nonumber\\
&=&\frac{1}{2S_{\bm q}}\int\frac{d^2q'}{4\pi}V_{\bm q'}\left(2\sin\left(\frac{1}{2}\bm q'\times\bm q\right)\right)^2\left(s_{\bm q'+\bm q}+s_{\bm q'-\bm q}-2s_{\bm q'}\right)\label{s4}
\end{eqnarray}
In the limit of small $|\bm q|$, we can expand Eq.(\ref{s4}) up to $O\left(|\bm q|^6\right)$ as follows:
\begin{eqnarray}
\lim_{|\bm q|\rightarrow 0}\delta E_{\bm q}&=&\frac{1}{2S_{\bm q}}\int\frac{d^2q'}{4\pi}V_{\bm q'}\left(2\sin\left(\frac{1}{2}\bm q'\times\bm q\right)\right)^2\left(q_aq_b\partial^a\partial^bs_{\bm q'}+\frac{1}{12}q_aq_bq_cq_d\partial^a\partial^b\partial^c\partial^ds_{\bm q'}+O\left(|\bm q|^6\right)\right)\nonumber\\
&=&\frac{1}{2S_{\bm q}}\int\frac{d^2q'}{4\pi}V_{\bm q'}\left(\epsilon^{ac}\epsilon^{bd}q'_aq'_bq_cq_d-\frac{1}{12}\epsilon^{ae}\epsilon^{bf}\epsilon^{cg}\epsilon^{dh}q'_aq'_bq'_cq'_dq_eq_fq_gq_h+O(|\bm q|^6)\right)\nonumber\\
&&\qquad\left(q_eq_f\partial^e\partial^fs_{\bm q'}+\frac{1}{12}q_eq_fq_gq_h\partial^e\partial^f\partial^g\partial^hs_{\bm q'}+O\left(|\bm q|^6\right)\right)\label{s5}
\end{eqnarray}
Repeated indices are summed over, and $\partial^a=\partial/\partial q_a$. The following two results are known:
\begin{eqnarray}
&&\lim_{|\bm q|\rightarrow 0}S_{\bm q}=\eta q^4+O(q^6)\\
&&s_{\bm q}=-\int\frac{d^2q'}{2\pi}e^{i\bm q\times \bm q'}s_{\bm q'}\label{s7}
\end{eqnarray}
and we take $q=|\bm q|$. These allow us to transform Eq.(\ref{s5}) into the following:
\begin{eqnarray}
\lim_{|\bm q|\rightarrow 0}\delta E_{\bm q}&=&\frac{1}{2S_{\bm q}}\int\frac{d^2q'd^2q''}{8\pi^2}\left(q_eq_f\epsilon^{eg}\epsilon^{fh}q''_gq''_hs_{\bm q''}-\frac{1}{12}q_eq_fq_gq_h\epsilon^{ek}\epsilon^{fl}\epsilon^{gm}\epsilon^{hn}q''_kq''_lq''_mq''_ns_{\bm q''}+O\left(|\bm q|^6\right)\right)\nonumber\\
&&\qquad\qquad\qquad\qquad \left(\epsilon^{ac}\epsilon^{bd}q'_aq'_bq_cq_d-\frac{1}{12}\epsilon^{ae}\epsilon^{bf}\epsilon^{cg}\epsilon^{dh}q'_aq'_bq'_cq'_dq_eq_fq_gq_h+O(|\bm q|^6)\right)V_{\bm q'}e^{i\bm q'\times\bm q''}\nonumber\\
&=&\frac{1}{2\eta}\int\frac{d^2q'd^2q''}{8\pi^2}V_{\bm q'}\left(\left(q'_x\right)^2\left(q''_x\right)^2-\frac{q^2}{12}q'^2_xq''^2_x\left(q'^2_x+q''^2_x\right)\right)s_{\bm q''}e^{i\bm q'\times\bm q''}+O\left(|\bm q|^4\right)\label{s8}
\end{eqnarray}
We have assumed rotational invariance here so that $V_{\bm q}=V_{|\bm q|}$, which gives $s_{\bm q}=s_{|\bm q|}$. The integration in Eq.(\ref{s8}) can be performed analytically, if we do the following expansion in the basis of the Laguerre polynomials:
\begin{eqnarray}
&&V_{|\bm q|}=\sum_nc_nL_n\left(q^2\right)e^{-\frac{q^2}{2}}\label{s9}\\
&&s_{|\bm q|}=\sum_nd_nL_n\left(q^2\right)e^{-\frac{q^2}{2}}\label{s10}
\end{eqnarray}
where $L_n\left(x\right)$ is the $n^{\text{th}}$ pseudopotential. Thus Eq.(\ref{s9}) is the usual pseudopotential expansion of the effective interaction, while Eq.(\ref{s10}) is possible because of Eq.(\ref{s7}). In both cases, $n$ can only be odd integers. We can thus rewrite Eq.(\ref{s8}) as follows:
\begin{eqnarray}
\lim_{|\bm q|\rightarrow 0}\delta E_{\bm q}&=&\frac{1}{256\eta}\Gamma^{mn}_{\left(1\right)}c_md_n+\frac{1}{768\eta}\Gamma^{mn}_{\left(2\right)}c_md_nq^2+O(q^4)\label{s11}\\
\Gamma^{mn}_{\left(1\right)}&=&16\int\frac{d^2q'd^2q''}{\pi^2}e^{-\frac{q'^2+q''^2}{2}}L_m\left(q'^2\right)L_n\left(q''^2\right)\left(q'_x\right)^2\left(q''_x\right)^2e^{i\bm q'\times\bm q''}\nonumber\\
&=&2\int q'q''dq'dq''e^{-\frac{q'^2+q''^2}{2}}L_m\left(q'^2\right)L_n\left(q''^2\right)q'^2q''^2\left(\text{BesselJ}\left(0,qq'\right)+\frac{2\cdot\text{BesselJ}\left(1,qq'\right)}{qq'}\right)\nonumber\\
&=&\frac{1}{2}\int dq'dq''e^{-\frac{q'+q''}{2}}L_m\left(q'\right)L_n\left(q''\right)q'q''\left(\prescript{}{0}{\mathbf{F}}_1\left(1,-\frac{q_1q_2}{4}\right)+\prescript{}{0}{\mathbf{F}}_1\left(2,-\frac{q_1q_2}{4}\right)\right)\nonumber\\
&=&[(1-m)m\delta_{m,2+n}-(1+m) (2+m)\delta_{m,n-2}+2(1+m+m^2)\delta_{m,n}]\left(-1\right)^m\\
\Gamma^{mn}_{\left(2\right)}&=&-4\int\frac{d^2q'd^2q''}{\pi^2}e^{-\frac{q'^2+q''^2}{2}}L_m\left(q'^2\right)L_n\left(q''^2\right)q'^2_xq''^2_x\left(q'^2_x+q''^2_x\right)e^{i\bm q'\times\bm q''}\nonumber\\
&=&-\int q'q''dq'dq''e^{-\frac{q'^2+q''^2}{2}}L_m\left(q'^2\right)L_n\left(q''^2\right)q'^2q''^2\left(q'^2+q''^2\right)\left(\text{BesselJ}\left(0,q'q''\right)+\frac{4\cdot\text{BesselJ}\left(1,q'q''\right)}{q'q''}\right)\nonumber\\
&=&-\frac{1}{4}\int dq'dq''e^{-\frac{q'+q''}{2}}L_m\left(q'\right)L_n\left(q''\right)q'q''\left(q'+q''\right)\left(\prescript{}{0}{\mathbf{F}}_1\left(1,-\frac{q_1q_2}{4}\right)+2\cdot\prescript{}{0}{\mathbf{F}}_1\left(2,-\frac{q_1q_2}{4}\right)\right)\nonumber\\
&=&\left(-1\right)^m[(2m-1)(m-1)m\delta_{m,2+n}+(1+m) (2+m)(2m+3)\delta_{m,n-2}-2(1+2m)(1+m+m^2)\delta_{m,n}]\qquad\qquad
\end{eqnarray}

\section{Relationship of neutral excitations to squeezed ground states}
Another family of trial wavefunctions can be defined as follows:
\begin{eqnarray}
|\xi_{\theta,\phi}\rangle=e^{i\alpha_{ab}\hat{\Lambda}^{ab}}|\psi_0\rangle,\qquad\hat{\Lambda}^{ab}=\frac{1}{4}\sum_i\{\hat R^a_i,\hat R^b_i\}
\end{eqnarray}
The algebra of the generator $\hat \Lambda^{ab}$ is given in the main text. The variational energy of the trial wavefunctions are given by:
\begin{eqnarray}
&&\delta E_\alpha=\langle\xi_{\theta,\phi}|\hat{\mathcal H}|\xi_{\theta,\phi}\rangle-E_0=\int\frac{d^2q}{4\pi}V_{\bm q}s_{\tilde{\bm q}}-E_0,\quad\tilde q_a=\left(\lambda^{-1}\right)^b_aq_b\label{s15}\\
&&s_{\tilde{\bm q}}=-\int\frac{d^2q'}{2\pi}s_{\bm q'}e^{i\tilde{\bm q}\times\bm q'}
\end{eqnarray}
The effect of the area-preserving deformation by $\hat U\left(\alpha\right)=e^{i\alpha_{ab}\hat\Lambda^{ab}}$ is to squeeze and rotate the metric by preserving the determinant of the metric. We thus have $\eta^{ab}\tilde q_a\tilde q_b=g^{ab}q_aq_b$ with the following metric:
\begin{eqnarray}
\eta=\left(\begin{array}{cc}
1& 0\\
0 & 1\end{array}\right)\qquad\qquad
g=\left(\begin{array}{cc}
\cosh\theta+\sinh\theta\cos\phi & \sinh\theta\sin\phi\\
\sinh\theta\sin\phi & \cosh\theta-\sinh\theta\cos\phi\end{array}\right)
\end{eqnarray}
Due to rotational invariance, we can take $\phi=0$. For small deformation, e.g. in the limit of $\theta\rightarrow 0$, we have $\tilde q_x=q_x+\frac{1}{2}\left(\theta+\frac{1}{4}\theta^2\right)q_x+O(\theta^3), \tilde q_y=q_y+\frac{1}{2}\left(-\theta+\frac{1}{4}\theta^2\right)q_y+O(\theta^3)$. Expansion of Eq.(\ref{s15}) leads to the following:
\begin{eqnarray}
\lim_{\theta\rightarrow 0}\delta E_\alpha&=&-\int \frac{d^2qd^2q'}{8\pi^2}V_{\bm q}s_{\bm q'}e^{i\bm q\times\bm q'}e^{\frac{i}{2}\left(\theta+\frac{1}{4}\theta^2\right)q_xq'_y+\frac{i}{2}\left(\theta-\frac{1}{4}\theta^2\right)q_yq'_x}-E_0\nonumber\\
&=&\frac{1}{8}\int \frac{d^2qd^2q'}{8\pi^2}V_{\bm q}s_{\bm q'}e^{i\bm q\times\bm q'}\left(q_x^2q'^2_y+q^2_yq'^2_x+2q_xq'_xq_yq'_y-i\left(q_xq_y'-q'_xq_y\right)\right)\theta^2+O(\theta^3)\nonumber\\
&=&\frac{\theta^2}{64}\Gamma^{mn}_{\left(1\right)}c_md_n+O(\theta^3)
\end{eqnarray}
One should note that the leading order of $\delta E_\alpha$ is $\theta^2$, so for any rotationally invariant Hamiltonian, the ground state variational energy is minimised when its intrinsic guiding center metric is undeformed. More importantly for small $\theta$, the variational energy gap of $|\xi_{\theta,\phi}\rangle$ is in the same form as Eq.(\ref{s11}), or the variational energy gap of the neutral excitations in the long wavelength limit.

\section{Nematic order of the neutral excitations}
The nematic order operator defined in the main text is given as follows:
\begin{eqnarray}
\hat{\mathcal N}&=&\int_0^{2\pi}\frac{d\theta}{2\pi}e^{2i\theta}\lim_{|\bm l|\rightarrow 0}\frac{1}{|\bm l|^2}\delta\hat\rho\left(\bm r+\bm l/2\right)\delta\hat\rho\left(\bm r-\bm l/2\right)\nonumber\\
&=&\int_0^{2\pi}\frac{d\theta}{2\pi}e^{2i\theta}\lim_{|\bm l|\rightarrow 0}\frac{1}{|\bm l|^2}\int\frac{d^2q_1d^2q_2}{4\pi^2}e^{i\bm r\cdot\left(\bm q_1+\bm q_2\right)}e^{\frac{i\bm l}{2}\left(\bm q_1-\bm q_2\right)}\delta\hat\rho_{\bm q_1}\delta\hat\rho_{\bm q_2}
\end{eqnarray}
Thus for translationally invariant systems, for any momentum eigenstate $|\psi_{\bm q}\rangle$, we have the following relationship:
\begin{eqnarray}
\langle\psi_{\bm q}|\hat{\mathcal N}|\psi_{\bm q'}\rangle=\int_0^{2\pi}\frac{d\theta}{2\pi}e^{2i\theta}\lim_{|\bm l|\rightarrow 0}\frac{1}{|\bm l|^2}\int\frac{d^2q_1}{2\pi}e^{i\bm r\cdot\left(\bm q-\bm q'\right)}e^{i\bm l\cdot \bm q_1}e^{i\frac{\bm l}{2}\left(\bm q'-\bm q\right)}\langle\psi_{\bm q}|\delta\hat\rho_{\bm q_1}\delta\hat\rho_{\bm q-\bm q'-\bm q_1}|\psi_{\bm q'}\rangle
\end{eqnarray}
With $\bm q=\bm q'$, the integration over the momentum amounts to the Fourier transform of the unregularised guiding center structure factor. The angle integration thus makes the nematic order vanish if $|\psi_{\bm q}\rangle$ is rotationally invariant (e.g. for ground state at $\bm q=0$). For neutral excitations that break rotational invariance, there could be a uniform nematic order given by:
\begin{eqnarray}
\langle\psi_{\bm q}|\hat{\mathcal N}|\psi_{\bm q}\rangle&=&\int_0^{2\pi}\frac{d\theta}{2\pi}e^{2i\theta}\lim_{|\bm l|\rightarrow 0}\frac{1}{|\bm l|^2}\int\frac{d^2q_1}{2\pi}e^{i\bm l\cdot\bm q_1}\langle\psi_{\bm q}|\delta\hat\rho_{\bm q_1}\delta\hat\rho_{-\bm q_1}|\psi_{\bm q}\rangle=-\int_0^{2\pi}\frac{d\theta}{2\pi}e^{2i\theta}\lim_{|\bm l|\rightarrow 0}\frac{1}{|\bm l|^2}s_{\bm q,\tilde{\bm l}}
\end{eqnarray}
where we have $s_{\bm q,\tilde{\bm l}}=\langle\psi_{\bm q}|\delta\hat\rho_{\tilde{\bm l}}\delta\hat\rho_{-\tilde{\bm l}}|\psi_{\bm q}\rangle-s_\infty, \tilde l_a=l_B^{-2}\epsilon_{ab}l^b$, and unregularised part of the structure factor vanishes with the angle integration. Thus the nematic order comes from the nematic properties of the state guiding center structure factor. We are also interested in the case of $\bm q'=-\bm q$, which gives us:
\begin{eqnarray}
\langle\psi_{\bm q}|\hat{\mathcal N}|\psi_{-\bm q}\rangle=e^{2i\bm r\cdot\bm q}\int_0^{2\pi}\frac{d\theta}{2\pi}e^{2i\theta}\lim_{|\bm l|\rightarrow 0}\frac{1}{|\bm l|^2}\int\frac{d^2q_1}{2\pi}e^{i\bm l\cdot \left(\bm q_1-\bm q\right)}\langle\psi_{\bm q}|\delta\hat\rho_{\bm q_1}\delta\hat\rho_{2\bm q-\bm q_1}|\psi_{-\bm q}\rangle
\end{eqnarray}
Thus for the nematic order of the neutral excitations defined in the main text, $|\psi^{\pm}_{\bm q}\rangle=\frac{1}{\sqrt{2S_{\bm q}}}\left(|\psi_{\bm q}\rangle\pm|\psi_{-\bm q}\rangle\right)$, we have the following:
\begin{eqnarray}
&&\langle\psi^{\pm}_{\bm q}|\hat{\mathcal N}|\psi^{\pm}_{\bm q}\rangle=\mathcal N_{q}^{\left(1\right)}\pm\cos 2qr\mathcal N_{q}^{\left(2\right)}\\
&&\mathcal N_{q}^{\left(1\right)}=-\frac{1}{S_{\bm q}}\int_0^{2\pi}\frac{d\theta}{2\pi}e^{2i\theta}\lim_{|\bm l|\rightarrow 0}\frac{1}{|\bm l|^2}s_{\bm q,\tilde{\bm l}}\\
&&\mathcal N_{q}^{\left(2\right)}=\frac{2}{S_{\bm q}}\int_0^{2\pi}\frac{d\theta}{2\pi}e^{2i\theta}\lim_{|\bm l|\rightarrow 0}\frac{1}{|\bm l|^2}\int\frac{d^2q_1}{2\pi}e^{i\bm l\cdot \left(\bm q_1-\bm q\right)}\langle\psi_{\bm q}|\delta\hat\rho_{\bm q_1}\delta\hat\rho_{2\bm q-\bm q_1}|\psi_{-\bm q}\rangle
\end{eqnarray}

\section{Model wavefunctions for two neutral excitations in $L=0$ sector}

For a single quadrupole or dipole excitation, the model wavefunction (for which the excitation is located at the north pole) is well known\cite{yangboss}. In the $L=0$ sector, two such excitations can be constructed by putting the two excitations at the north and south pole respectively. For two quadrupole excitations, the root configuration of the model wavefunction can be given as follows:
\begin{eqnarray}\label{sroot1}
\d{1}\d{1}0\textsubring{0}\textsubring{0}00100100\cdots 10010\textsubring{0}\textsubring{0}0\d{1}\d{1}
\end{eqnarray}
where the solid and open circles indicate the locations of quasiparticles (of charge $e/3$) and quasiholes (of charge $-e/3$). The basis of the model wavefunction only involves monomials or Slater determinants squeezed from the root configuration in Eq.(\ref{sroot1}). For model wavefunctions containing two dipole excitations in the $L=0$ sector, the quasiholes will be separated as far from the quasielectrons as possible. For example with $N_e=12, N_o=34$, the root configuration for such a state is given by:
\begin{eqnarray}\label{sroot2}
\d{1}\d{1}0\textsubring{0}010010010010\textsubring{0}\textsubring{0}010010010010\textsubring{0}0\d{1}\d{1}
\end{eqnarray}
The microscopic model wavefunctions can be obtained from Eq.(\ref{sroot1}) or Eq.(\ref{sroot2}) using the method analogous to Ref.\cite{yangboss}. For example from Eq.(\ref{sroot1}), we remove the quasielectrons at the north and south pole from the root configuration to obtain:
\begin{eqnarray}
000\textsubring{0}\textsubring{0}00100100\cdots 10010\textsubring{0}\textsubring{0}0000
\end{eqnarray}
We then require this root configuration to give the Jack polynomial that is the zero energy state of the $V_1$ pseudopotential. Formally, let $|\psi_{\text{2-quadrupole}}\rangle$ be the model wavefunction with Eq.(\ref{sroot1}) as the root configuration and consisting of only basis squeezed from the root configuration. We then require:
\begin{eqnarray}
&&\hat V_1\hat c_1\hat c_2\hat c'_1\hat c'_2|\psi_{\text{2-quadrupole}}\rangle=0\label{sv1}\\
&&\hat L^{+}|\psi_{\text{2-quadrupole}}\rangle=0\label{shw}
\end{eqnarray}
In Eq.(\ref{sv1}), $\hat c_1, \hat c_2$ are annihilation operators removing the two electrons at the north pole, while $\hat c'_1, \hat c'_2$ are annihilation operators removing the two electrons at the south pole; $\hat V_1$ is the two-body interaction Hamiltonian with $V_1$ pseudopotential. Eq.(\ref{shw}) just imposes the highest weight condition to $|\psi_{\text{2-quadrupole}}\rangle$. Combining both Eq.(\ref{sv1}) and Eq.(\ref{shw}) will lead to a unique state which is the model two-quadrupole state in the $L=0$ sector. One can also diagonalise the respective squeezed basis with $V_1$ pseudopotential interaction, and the model wavefunctions are obtained as the ground state. In the main text, since we are not required to go to very large system sizes, we employ the latter method to obtain the model states and the overlaps.

\end{widetext}

\end{document}